\documentclass[aps, prd, onecolumn, tightenlines, notitlepage, superscriptaddress, nofootinbib, preprintnumbers, floatfix,showkeys,11pt,altaffilletter]{revtex4-2}

\usepackage[official]{eurosym}
\usepackage[normalem]{ulem}
\usepackage{amstext}
\usepackage{graphicx}
\graphicspath{{figures/}}
\usepackage{url}
\usepackage{color}
\usepackage{ulem}
\usepackage[version=4]{mhchem}
\usepackage[utf8]{inputenc}
\usepackage{fontawesome}
\usepackage{yfonts}

\usepackage{epsfig,amsfonts,mathrsfs,amsmath,amssymb,graphicx,color,slashed,multirow}
\usepackage{amsmath,latexsym,amssymb,graphicx,slashed,color,enumerate,url,cancel,gensymb}
\usepackage{nccmath} 
\usepackage{textcomp}

\allowdisplaybreaks 

\usepackage[x11names]{xcolor}
\usepackage[colorlinks,pdfstartview=FitV,breaklinks=true]{hyperref}
\usepackage{adjustbox}
\usepackage{textgreek} 
\usepackage{caption, subcaption} 
\usepackage{ragged2e} 
\DeclareCaptionJustification{justified}{\justifying}
\usepackage{booktabs} 
\usepackage{float}
\usepackage{orcidlink}
\usepackage{lmodern}
\usepackage{ae,aecompl}
\usepackage{appendix}

\hypersetup{colorlinks,citecolor= nicered,linkcolor= blue}
\definecolor{nicered}{rgb}{0.7,0.1,0.1}
\definecolor{nicegreen}{rgb}{0.1,0.5,0.1}
\makeatletter
    \newcommand{\colorboxed}[3][white]{\fcolorbox{#2}{#1}{\m@th$\displaystyle#3$}}
\makeatother

\def\d{\mathrm{d}}
\def\cevns{CE$\nu$NS}
\def\eves{E$\nu$ES}
\def\d{\mathrm{d}}

\definecolor{blue(ncs)}{rgb}{0.0, 0.53, 0.74}
\definecolor{chromeyellow}{rgb}{1.0, 0.56, 0.0}
\definecolor{amber(sae/ece)}{rgb}{1.0, 0.49, 0.0}
\definecolor{regalia}{rgb}{0.32, 0.18, 0.5}

\newcommand{\qtransfer}{\left|\mathbf{q}\right|}

\AtBeginDocument{\hypersetup{citecolor=regalia,linkcolor=regalia,urlcolor=regalia}}
\usepackage{appendix}

\begin{document}
\title{{\LARGE Up-scattering production of a sterile fermion at DUNE: complementarity with spallation
source and direct detection experiments}}
\author{Pablo M. Candela~\orcidlink{0009-0009-8416-9295}}\email{pamuca@ific.uv.es}
\affiliation{Instituto de F\'{i}sica Corpuscular (CSIC-Universitat de Val\`{e}ncia), Parc Cient\'ific UV C/ Catedr\'atico Jos\'e Beltr\'an, 2 E-46980 Paterna (Valencia) - Spain}
\author{Valentina De Romeri~\orcidlink{0000-0003-3585-7437}}\email{deromeri@ific.uv.es}
\affiliation{Instituto de F\'{i}sica Corpuscular (CSIC-Universitat de Val\`{e}ncia), Parc Cient\'ific UV C/ Catedr\'atico Jos\'e Beltr\'an, 2 E-46980 Paterna (Valencia) - Spain}
\author{Pantelis Melas~\orcidlink{0009-0000-2901-3444}}\email{pmelas@fnal.gov}
\affiliation{Department of Physics, National and Kapodistrian University
of Athens, Zografou Campus GR-15772 Athens, Greece}
\author{Dimitrios K. Papoulias~\orcidlink{0000-0003-0453-8492}}\email{dkpapoulias@phys.uoa.gr}
\affiliation{Department of Physics, National and Kapodistrian University
of Athens, Zografou Campus GR-15772 Athens, Greece}
\author{Niki Saoulidou~\orcidlink{0000-0001-6958-4196}}\email{Niki.Saoulidou@cern.ch}
\affiliation{Department of Physics, National and Kapodistrian University
of Athens, Zografou Campus GR-15772 Athens, Greece}

\keywords{neutrinos,  COHERENT, LZ, XENONnT, DUNE, light mediators, sterile fermion}

\begin{abstract}
We investigate the possible production of a MeV-scale sterile fermion through the up-scattering of neutrinos on nuclei and atomic electrons at different facilities. We consider a phenomenological model that adds a new fermion to the particle content of the Standard Model and we allow for all possible Lorentz-invariant non-derivative interactions (scalar, pseudoscalar, vector, axial-vector and tensor) of neutrinos
with electrons and first-generation quarks. We first explore the sensitivity of the DUNE experiment to this scenario, by simulating elastic neutrino-electron scattering events in the near detector. We consider both options of a standard and a tau-optimized neutrino beams, and investigate  the impact of a mobile detector that can be moved off-axis with respect to the beam. Next, we infer constraints on the typical coupling, new fermion and mediator masses from elastic neutrino-electron scattering events induced by solar neutrinos in two current dark matter direct detection experiments, XENONnT and LZ. Under the assumption that the new mediators couple also to first-generation quarks, we further set constraints on the up-scattering production of the sterile fermion using coherent elastic neutrino-nucleus scattering data from the COHERENT experiment. Moreover, we set additional constraints  assuming that the sterile fermion may decay within the detector. We finally compare our results and discuss how  these facilities are sensitive to different regions of the relevant parameter space due to kinematics arguments and can hence provide complementary information on the up-scattering production of a sterile fermion.

\end{abstract}
\maketitle

\section{Introduction}
The experimental evidence for neutrino oscillations~\cite{deSalas:2020pgw,Esteban:2020cvm} provides strong motivation for the existence of new physics beyond the Standard Model (SM). On the theoretical side, many SM extensions meant to accommodate neutrino masses call upon the introduction of new sterile states that mix with active neutrinos (see e.g.~\cite{Mohapatra:2005wg}). Such sterile neutrinos, whose masses can in principle range over many orders of magnitude depending on the
actual mechanism that originates neutrino masses, have received increasing attention in the recent years. Despite their extended phenomenological implications, comprising cosmological and astrophysical ones, no positive evidence of heavy sterile neutrinos   has been found to date~\cite{Atre:2009rg,Drewes:2015iva,Abdullahi:2022jlv}. Nevertheless, searches for new sterile fermions  (and in particular leptons) in the MeV-GeV range are theoretically motivated and particularly appealing, as they are in principle accessible at a range of laboratory experiments, including high-intensity and high-energy facilities (see~\cite{Abdullahi:2022jlv} for a recent review).

A crucial role in these searches may be played by neutrino oscillation experiments, performing beam-dump style measurements~\cite{Arguelles:2019xgp}. The Deep Underground Neutrino Experiment (DUNE) is a next-generation long-baseline neutrino oscillation experiment with manifold applications besides its primary goal of a precise determination of neutrino oscillation patterns~\cite{DUNE:2020lwj,DUNE:2020ypp,DeRomeri:2016qwo}. Its Near Detector (ND) complex, located at the Fermi National Accelerator Laboratory (FNAL), is characterized by strong particle reconstruction capabilities typical of Liquid Argon (LAr) detectors, which, together with a high-power, broadband neutrino beam anticipate a great potential in searches for new physics~\cite{DUNE:2021tad,DUNE:2020fgq}. The ND will be located approximately 574 m from the neutrino source and will consist of three primary detector components with the capability for two of them to move off the beam axis~\cite{DUNE:2021tad}. This feature is called the DUNE Precision Reaction-Independent Spectrum Measurement (PRISM) and is primarily designed to produce data-driven predictions of the oscillated neutrino event rate spectrum at the far detector~\cite{Hasnip:2023ygr,DUNE:2021tad}, and to help constraining neutrino cross sections and flux-related uncertainties~\cite{Marshall:2019vdy}. The  liquid argon time-projection chamber (LArTPC) is expected to have a 67 tonne fiducial volume and to be mobile up to 30.5 m off-axis~\cite{DUNE:2021tad}. With these characteristics, the DUNE-ND --- specially if equipped with the off-axis option --- will offer excellent opportunities to search for low-energy new physics~\cite{Caratelli:2022llt}, for instance in the form of heavy neutral leptons~\cite{Berryman:2019dme, Ballett:2019bgd,Coloma:2020lgy,Breitbach:2021gvv} or light DM~\cite{DeRomeri:2019kic,Batell:2022xau,Brdar:2022vum}.

Besides ``traditional" neutrino oscillation experiments~\cite{Kajita:2016cak,McDonald:2016ixn}, the recent observation of coherent elastic neutrino-nucleus scattering (\cevns)~\cite{COHERENT:2017ipa,COHERENT:2020iec,COHERENT:2021xmm} has proven to be a powerful tool to probe the neutrino sector and search for new physics~\cite{Barranco:2005yy,Abdullah:2022zue,Papoulias:2019xaw, DeRomeri:2022twg}. Experiments using neutrinos from stopped-pion beams, like COHERENT~\cite{Akimov:2022oyb},  Coherent CAPTAIN-Mills~\cite{CCM:2021leg} and the planned European Spallation Source~\cite{Baxter:2019mcx} are specially suited to study new physics in the form of neutrino non-standard interactions (NSI)~\cite{Chatterjee:2022mmu} or neutrino generalized interactions (NGI)~\cite{Lindner:2016wff,Rodejohann:2017vup,AristizabalSierra:2018eqm,DeRomeri:2022twg}. These experiments can also probe the existence of new light~\cite{Kosmas:2017zbh,Blanco:2019vyp,Miranda:2020syh,Miranda:2021kre,DeRomeri:2022twg} or heavy states~\cite{Calabrese:2022mnp,DeRomeri:2023cjt}, the inelastic transition into a new, heavy, sterile neutrino through the presence of a 
active-sterile transition magnetic moment~\cite{McKeen:2010rx,Miranda:2021kre,Bolton:2021pey,DeRomeri:2022twg} or the up-scattering production of a new fermion at the MeV scale through new scalar or vector mediators~\cite{Brdar:2018qqj,Chang:2020jwl,Chao:2021bvq,Candela:2023rvt,Alonso-Gonzalez:2023tgm}.

Apace with neutrino experiments, there is another vast array of facilities that have currently reached impressive sensitivities: dark matter (DM) direct detection (DD) experiments~\cite{Schumann:2019eaa,Baxter:2021pqo}. These underground detectors looking for ghostly particles of cosmological or astrophysical origin share many common features with neutrino experiments. Despite their searches for DM in the form of weakly interacting massive particles (WIMPs) have given null results till now~\cite{Billard:2021uyg}, the incredibly low thresholds, large volumes and background reduction of these facilities allow them to observe astrophysical neutrinos. While constituting an irreducible background for WIMP searches~\cite{Monroe:2007xp,Vergados:2008jp,Strigari:2009bq,Billard:2013qya,OHare:2016pjy,OHare:2021utq}, the observation of solar neutrinos at DM DD experiments may provide new unique opportunities to improve the neutrino physics picture~\cite{Cerdeno:2016sfi,Dutta:2017nht,Gelmini:2018gqa,Essig:2018tss, Amaral:2020tga,Dutta:2020che, Amaral:2021rzw,deGouvea:2021ymm,AtzoriCorona:2022jeb,Giunti:2023yha,DeRomeri:2024dbv}. The latest generation of  dual-phase liquid xenon (LXe) detectors like those employed by the XENON~\cite{XENON:2017lvq,XENON:2018voc,XENON:2022ltv} and LUX-ZEPLIN (LZ)~\cite{LZ:2015kxe,LZ:2019sgr,LZ:2022lsv,LZ:2023poo} experiments, have reached world-leading sensitivities not only on WIMPs but also on new-physics scenarios beyond the SM.  
Solar (and other sources, e.g., atmospheric and diffuse supernova background) neutrinos can induce not only \cevns~but also elastic neutrino-electron scattering (E$\nu$ES) events in DM DD experiments, both having a strong degeneracy with the corresponding DM-electron~\cite{Essig:2018tss,Wyenberg:2018eyv,Carew:2023qrj} and DM-nucleus signals~\cite{Billard:2013qya,OHare:2021utq,AristizabalSierra:2021kht}. 
However, due to energy-threshold limitations, solar neutrinos at DM DD experiments are currently detected mainly through E$\nu$ES~\cite{Majumdar:2021vdw,AtzoriCorona:2022jeb,Khan:2022bel,Amaral:2023tbs,DeRomeri:2024dbv}.

In this paper we explore the complementarity of different facilities in probing the up-scattering production of a new sterile fermion $\chi$ with a mass in the MeV range. We consider a simple, phenomenological extension of the SM with the addition of $\chi$, a generic sterile fermion, and new neutrino-quark and neutrino-electron interactions (scalar, pseudoscalar, vector, axial-vector or tensor). In this regard, we extend the NGI scenario~\cite{Lee:1956qn,Lindner:2016wff,AristizabalSierra:2018eqm,DeRomeri:2022twg,Melas:2023olz} by allowing for the inelastic transition of an active neutrino into a new fermion. For the sake of completeness, we work in a light mediator regime thus exploring a wider region of parameter space. 
We are mainly motivated  to explore the sensitivity to this scenario through \eves~events at the DUNE-ND. However, in order to provide a more complete picture, we also analyze current data from DM DD experiments, namely XENONnT~\cite{XENON:2022ltv} and LZ~\cite{LZ:2022lsv}, hence testing the up-scattering production of $\chi$ in \eves~events induced by solar (mainly $pp$ and $^{7}$Be) neutrinos. Moreover, and under the assumption that the new mediator  couples  not only to electrons but also to quarks, we investigate the up-scattering production of $\chi$ through \cevns~by analyzing the most recent COHERENT CsI dataset~\cite{COHERENT:2021xmm}. Given the different neutrino sources, solar neutrinos for XENONnT and LZ, neutrinos from pions decaying at rest for COHERENT and a proton beam for DUNE, these facilities will be able to explore different regions in parameter space, being sensitive to different new fermion masses.

One similar framework is the \textit{dipole portal}, that consists in the
possible production of a heavy neutral lepton due to the presence of active-sterile neutrino transition
magnetic moments~\cite{Magill:2018jla,Brdar:2020quo,Vogel:1989iv,Balantekin:2013sda}. 
The implications of this scenario have been studied for DUNE~\cite{Schwetz:2020xra,Atkinson:2021rnp,Ovchynnikov:2022rqj}, COHERENT~\cite{DeRomeri:2022twg,Bolton:2021pey,Miranda:2021kre} and DM DD experiments~\cite{Shoemaker:2018vii,Alonso-Gonzalez:2023tgm,Shoemaker:2020kji}. Here, we go one step beyond this scenario by considering not electromagnetic-type but generalized interactions that may lead to the up-scattering production of $\chi$. Searches for new  generalized interactions have also been extensively done in the literature, for example at DUNE~\cite{Bischer:2018zcz,Chakraborty:2021apc,Melas:2023olz}, neutrino-electron scattering experiments~\cite{Bilmis:2015lja,Chen:2021uuw}, DM DD experiments~\cite{Schwemberger:2022fjl,Demirci:2023tui} and \cevns~experiments~\cite{AristizabalSierra:2018eqm,Lindner:2016wff}.

Previous searches involving neutrino up-scattering into a sterile fermion exist in the literature~\cite{Brdar:2018qqj,Chang:2020jwl,Chao:2021bvq,Candela:2023rvt, Chen:2021uuw}.  Reference~\cite{Brdar:2018qqj} explored up-scattering of neutrinos into a sterile fermion, but assuming only a scalar mediator. Moreover, it exclusively focused on CE$\nu$NS data, relying on the first (old) COHERENT-CsI dataset and providing projected sensitivities for CONUS. Refs.~\cite{Chang:2020jwl,Chao:2021bvq,Chen:2021uuw} analyzed neutrino up-scattering into a sterile fermion via CE$\nu$NS using the old COHERENT-CsI and LAr data, and via E$\nu$ES using TEXONO, CHARM and XENON1T data. They assumed effective interactions by integrating out the explicit dependence on the mediators, unlike our present work. In Ref.~\cite{Candela:2023rvt} a combined analysis of CE$\nu$NS plus E$\nu$ES events using the most recent COHERENT-CsI and LAr data and assuming light mediators was presented. However, the latter analysis was restricted to vector and scalar interactions only. In the present work, we build upon Ref.~\cite{Candela:2023rvt} and expand the COHERENT analysis by considering all Lorentz-invariant interactions with explicit mediator dependence. Additionally, we set new limits on the up-scattering of neutrinos through generalized light mediators using recent E$\nu$ES measured data induced by solar neutrinos at XENONnT and LZ. We further provide projected sensitivities to this scenario via an E$\nu$ES-based analysis at the DUNE-ND using simulated data. Finally we quantify constraints coming out from the subsequent decays of $\chi$.\\

The remainder of the work is organized as follows: in Section~\ref{sec:dark-fermion-production} we discuss the theoretical scenario under consideration and the relevant \eves~and \cevns~cross sections for the new fermion production. In Section~\ref{sec:data-analysis} we present the details of the analyses of current DM DD and COHERENT data, together with the simulation of DUNE sensitivity. In Section~\ref{sec:results} we show the results of our analyses and we present our final discussion in Section~\ref{sec:conc}.

\section{Neutrino up-scattering into a sterile fermion}\label{sec:dark-fermion-production}

We consider the possible production of a sterile fermion (SF) $\chi$ from the up-scattering of neutrinos on atomic electrons or nuclei:
\begin{equation} \label{eq:df-process-production}
    \nu_\ell e \rightarrow \chi e, \qquad \nu_\ell \mathcal{N} \rightarrow \chi \mathcal{N},
\end{equation}
where $\nu_\ell$  is a neutrino or antineutrino of any flavor $\ell=e,\mu,\tau$, while the nature of the sterile fermion is that of a spin-1/2 field. In the case of an incoming antineutrino, the corresponding produced SF is its antiparticle.

Working in the framework of a simplified phenomenological scenario without invoking any concrete UV-complete model, we extend the SM Lagrangian at energies below the electroweak scale with 
\begin{equation} \label{eq:effective-lagrangian}
     \mathcal{L}_{\mathrm{SF}}^{a} \supseteq \dfrac{G_F}{\sqrt{2}}\, \varepsilon_\ell^a \left(\overline{\chi}\, \Gamma^a P_L \,\nu_\ell\right)\left(\overline{f}\, \Gamma_a  f\right) + \mathrm{H.c.}
\end{equation}
This Lagrangian encodes new neutral-current neutrino-$\chi$ interactions via one of the following generalized interactions: scalar ($S$), pseudoscalar ($P$), vector ($V$), axial-vector ($A$) or tensor ($T$). These are summarized through the subscript $a \equiv \left\{S,\, P,\, V,\, A,\, T\right\}$ as $\Gamma^a \equiv \left\{I,\, i\gamma^5,\, \gamma^\mu,\, \gamma^\mu \gamma^5,\, \sigma^{\mu\nu}\right\}$, with $\sigma^{\mu \nu} \equiv \frac{i}{2} [\gamma^\mu, \gamma^\nu]$. Here, $G_F$ is the Fermi coupling constant, $\varepsilon_\ell^a$ is a coupling that quantifies the strength of the interaction of type $a$ for a neutrino $\nu_\ell$, $\chi$ is the SF field, $P_L \equiv (1 - \gamma^5) / 2$ is the left-handed projector and $f$ is the fermion field that corresponds to electrons in \eves~or up and down quarks in  \cevns, i.e., $f = u,\, d,\, e^-$.

Given the Lagrangian~\eqref{eq:effective-lagrangian}, one can compute the cross section for each interaction in terms of the target's recoil energy.  Concerning the latter, DM DD experiments like  LZ and XENONnT involve a  typical momentum transfer of $\mathcal{O}$(100)~keV, while for COHERENT it is of the order of  $\mathcal{O}$(10)~MeV and finally for the case of DUNE it is in the range of a few GeV. The magnitude of the three-momentum transfer is defined as $|\mathbf{q}| \approx \sqrt{2 m_e T_e}$ for scattering on electrons or $|\mathbf{q}| \approx \sqrt{2 m_\mathcal{N} T_\mathcal{N}}$ for scattering on nuclei. It is not unreasonable to think about a light mediator with a mass comparable to the typical momentum transfer of these experiments. Thus, we  extend our discussion by including the possibility of having light mediators and we translate the effective-interaction couplings through the following substitution
\begin{equation}\label{eq:effective-light-mediator-transformation}
    G_F^2 |\varepsilon_\ell^a|^2 \rightarrow \dfrac{2 g_a^4}{(m_a^2 + |\mathbf{q}|^2)^2} ,
\end{equation}
where $m_a$ is the mass of the mediator of type $a \equiv \left\{S,\, P,\, V,\, A,\, T\right\}$ and $g_a$ is the new coupling that quantifies the strength of the interaction. In Eq.~\eqref{eq:effective-light-mediator-transformation} we have re-scaled the coupling constant and we have explicitly indicated the propagator dependence on the mediator mass. In addition, we have assumed that the coupling $\varepsilon_\ell^a$ is the same for every neutrino flavor $\nu_\ell$.  After this transformation, the resulting expressions for the \eves~cross section are
\begin{align}
    \left.\dfrac{\d \sigma_{\nu_\ell \mathcal{A}}}{\d T_e}\right|_\mathrm{\mathrm{E\nu ES}}^{\mathrm{S}} (E_\nu, T_e) =& \, \dfrac{m_e g_S^4}{4\pi (m_S^2 + 2m_e T_e)^2} Z_{\mathrm{eff}}^{\mathcal{A}}\left(T_e\right) \left(1 + \dfrac{T_e}{2 m_e}\right) \left(\dfrac{m_e T_e}{E_\nu^2} + \dfrac{m_\chi^2}{2 E_\nu^2}\right), \label{eq:cross-section-scalar-ES} \\[4pt]
    \left.\dfrac{\d \sigma_{\nu_\ell \mathcal{A}}}{\d T_e}\right|_\mathrm{\mathrm{E\nu ES}}^\mathrm{P} (E_\nu, T_e) =& \, \dfrac{m_e g_P^4}{4\pi (m_P^2 + 2m_e T_e)^2} Z_{\mathrm{eff}}^{\mathcal{A}}\left(T_e\right) \dfrac{T_e}{2 m_e} \left(\dfrac{m_e T_e}{E_\nu^2} + \dfrac{m_\chi^2}{2 E_\nu^2}\right), \label{eq:cross-section-pseudoscalar-ES} \\[4pt]
    \left.\dfrac{\d \sigma_{\nu_\ell \mathcal{A}}}{\d T_e}\right|_\mathrm{\mathrm{E\nu ES}}^\mathrm{V} (E_\nu, T_e) =& \, \dfrac{m_e g_V^4}{2\pi (m_V^2 + 2m_e T_e)^2} Z_{\mathrm{eff}}^{\mathcal{A}}\left(T_e\right) \nonumber \\[2pt]
    &\times \left[\left(1 - \dfrac{m_e T_e}{2 E_\nu^2} - \dfrac{T_e}{E_\nu} + \dfrac{T_e^2}{2 E_\nu^2}\right) - \dfrac{m_\chi^2}{4 E_\nu^2}\left(1 + \dfrac{2 E_\nu}{m_e} - \dfrac{T_e}{m_e}\right)\right], \label{eq:cross-section-vector-ES} \\[4pt]
    \left.\dfrac{\d \sigma_{\nu_\ell \mathcal{A}}}{\d T_e}\right|_\mathrm{\mathrm{E\nu ES}}^\mathrm{A} (E_\nu, T_e) =& \, \dfrac{m_e g_A^4}{2\pi (m_A^2 + 2m_e T_e)^2} Z_{\mathrm{eff}}^{\mathcal{A}}\left(T_e\right) \nonumber \\[2pt]
    &\times \left[\left(1 + \dfrac{m_e T_e}{2 E_\nu^2} - \dfrac{T_e}{E_\nu} + \dfrac{T_e^2}{2 E_\nu^2}\right) + \dfrac{m_\chi^2}{4 E_\nu^2}\left(1 - \dfrac{2 E_\nu}{m_e} + \dfrac{T_e}{m_e}\right)\right], \label{eq:cross-section-axialvector-ES} \\[4pt]
    \left.\dfrac{\d \sigma_{\nu_\ell \mathcal{A}}}{\d T_e}\right|_\mathrm{\mathrm{E\nu ES}}^\mathrm{T} (E_\nu, T_e) =& \, \dfrac{4 m_e g_T^4}{\pi (m_T^2 + 2m_e T_e)} Z_{\mathrm{eff}}^{\mathcal{A}}\left(T_e\right) \nonumber \\[2pt]
    &\times \left[\left(1 - \dfrac{m_e T_e}{4 E_\nu^2} - \dfrac{T_e}{E_\nu} + \dfrac{T_e^2}{4 E_\nu^2}\right) - \dfrac{m_\chi^2}{4 E_\nu^2}\left(\dfrac{1}{2} + \dfrac{2 E_\nu}{m_e} - \dfrac{T_e}{2 m_e}\right)\right], \label{eq:cross-section-tensor-ES}
\end{align}
where $T_e$ is the electron recoil energy, $E_\nu$ is the incident neutrino energy, $m_e$ is the electron mass and $m_\chi$ is the SF mass. Notably, the results for the vector and scalar cases are consistent with the cross sections presented in Ref.~\cite{Candela:2023rvt}  (and~\cite{Chen:2021uuw} in the regime of effective interactions), while in the limit of $m_\chi \to 0$ the NGI \eves~cross sections presented in Ref.~\cite{Melas:2023olz} are recovered. Tabulated values of  the effective charges $Z_{\mathrm{eff}}^{\mathcal{A}}\left(T_e\right)$,  indicating the amount of atomic electrons that can be ionized for a given energy deposition $T_e$ and different isotopes, are given in Appendix \ref{sec:appendix-effective-charge}.  Let us clarify that, based upon the NGI definition  given in  Lagrangian~\eqref{eq:effective-lagrangian} that denotes all possible four-fermion non-derivative Lorentz structures, we have included the tensor case also in the light-mediator regime, in the spirit of performing the most comprehensive phenomenological analysis.

In addition to the study of \eves, we consider the production of the SF via up-scattering on nuclei, since  we are interested in the recent COHERENT data, involving \cevns~events. To obtain these cross sections, one can replace in Eqs.~(\ref{eq:cross-section-scalar-ES}--\ref{eq:cross-section-tensor-ES}) the electron mass with the nuclear mass, $m_e \rightarrow m_\mathcal{N}$, the effective charge with the weak  nuclear form factor, $Z_{\mathrm{eff}}^{\mathcal{A}}\left(T_e\right) \rightarrow F_W^2(\qtransfer^2)$, and the quark-level couplings $g_a$ with the nuclear-level ones, $C_a$,  through the typical procedure also adopted in DM DD searches and described in~\cite{Cirelli:2013ufw}. 
Following these steps, we obtain
\begin{align}
    \left.\dfrac{\d \sigma_{\nu_\ell \mathcal{N}}}{\d T_\mathcal{N}}\right|_\mathrm{CE \nu NS}^\mathrm{S} (E_\nu, T_\mathcal{N}) =& \, \dfrac{m_\mathcal{N} C_S^4}{4\pi (m_S^2 + 2m_\mathcal{N} T_\mathcal{N})^2} F_W^2(\qtransfer^2) \left(1 + \dfrac{T_\mathcal{N}}{2 m_\mathcal{N}}\right) \left(\dfrac{m_\mathcal{N} T_\mathcal{N}}{E_\nu^2} + \dfrac{m_\chi^2}{2 E_\nu^2}\right), \label{eq:cross-section-scalar-CEvNS} \\[4pt]
    \left.\dfrac{\d \sigma_{\nu_\ell \mathcal{N}}}{\d T_\mathcal{N}}\right|_\mathrm{CE \nu NS}^\mathrm{P} (E_\nu, T_\mathcal{N}) =& \, \dfrac{m_\mathcal{N} C_P^4}{4\pi (m_P^2 + 2m_\mathcal{N} T_\mathcal{N})^2} F_W^2(\qtransfer^2) \dfrac{T_\mathcal{N}}{2 m_\mathcal{N}} \left(\dfrac{m_\mathcal{N} T_\mathcal{N}}{E_\nu^2} + \dfrac{m_\chi^2}{2 E_\nu^2}\right), \label{eq:cross-section-pseudoscalar-CEvNS} \\[4pt]
    \left.\dfrac{\d \sigma_{\nu_\ell \mathcal{N}}}{\d T_\mathcal{N}}\right|_\mathrm{CE \nu NS}^\mathrm{V} (E_\nu, T_\mathcal{N}) =& \, \dfrac{m_\mathcal{N} C_V^4}{2\pi (m_V^2 + 2m_\mathcal{N} T_\mathcal{N})^2} F_W^2(\qtransfer^2) \nonumber \\[4pt]
    &\times \left[
        \left(1 - \dfrac{m_\mathcal{N} T_\mathcal{N}}{2 E_\nu^2} - \dfrac{T_\mathcal{N}}{E_\nu} + \dfrac{T_\mathcal{N}^2}{2 E_\nu^2}\right) - \dfrac{m_\chi^2}{4 E_\nu^2}\left(1 + \dfrac{2 E_\nu}{m_\mathcal{N}} - \dfrac{T_\mathcal{N}}{m_\mathcal{N}}\right)
    \right], \label{eq:cross-section-vector-CEvNS} \\[4pt]
    \left.\dfrac{\d \sigma_{\nu_\ell \mathcal{N}}}{\d T_\mathcal{N}}\right|_\mathrm{CE \nu NS}^\mathrm{A} (E_\nu, T_\mathcal{N}) =& \,  \dfrac{2 m_\mathcal{N}}{2J + 1} \dfrac{g_A^4}{(m_A^2 + 2 m_\mathcal{N}  T_\mathcal{N})^2} \nonumber \\[4pt]
    &\hspace{-8em}\medmath{\times \Bigg\lbrace \left[
        \left(
            2 + \dfrac{m_\mathcal{N}  T_\mathcal{N}}{E_\nu^2} - \dfrac{2 T_\mathcal{N}}{E_\nu} -  \dfrac{T_\mathcal{N}}{m_\mathcal{N}} + \dfrac{T_\mathcal{N}^2}{2 E_\nu^2} + \dfrac{T_\mathcal{N}^2}{m_\mathcal{N} E_\nu }
        \right)  - \dfrac{m_\chi^2}{2 E_\nu^2} \left(
            1 + \dfrac{3 E_\nu}{m_\mathcal{N}} + \dfrac{m_\chi^2 }{m_\mathcal{N} T_\mathcal{N}} - \dfrac{ T_\mathcal{N}}{2 m_\mathcal{N}}
        \right)
    \right] \tilde{S}^{\mathcal{T}}(\qtransfer^2)} \nonumber \\[4pt]
    &\hspace{-7em} \medmath{+ 2\, \left[ \dfrac{T_\mathcal{N}}{E_\nu}\left(
        \dfrac{E_\nu}{m_\mathcal{N}} - \dfrac{T_\mathcal{N}}{2 E_\nu} - \dfrac{T_\mathcal{N}}{m_\mathcal{N}} \vphantom{\dfrac{m_\chi^2}{m_\mathcal{N} T_\mathcal{N}}}
    \right) + \dfrac{m_\chi^2}{2 E_\nu^2} \left(
        2 + \dfrac{3 E_\nu}{m_\mathcal{N}} + \dfrac{m_\chi^2}{m_\mathcal{N} T_\mathcal{N}} - \dfrac{T_\mathcal{N}}{2 m_\mathcal{N}}
    \right)\right] \tilde{S}^\mathcal{L}(\qtransfer^2) \Bigg\rbrace} \, , \label{eq:cross-section-axialvector-CEvNS} \\[4pt]
   \left.\dfrac{\d \sigma_{\nu_\ell \mathcal{N}}}{\d T_\mathcal{N}}\right|_\mathrm{CE \nu NS}^\mathrm{T} (E_\nu, T_\mathcal{N}) =& \, \dfrac{m_\mathcal{N} }{2J + 1} \dfrac{g_T^4}{(m_T^2 + 2 m_\mathcal{N} T_\mathcal{N})^2} \nonumber \\[4pt]
    & \hspace{-8em} \medmath{\times \Bigg\lbrace \left[\left(2 - \dfrac{m_\mathcal{N} T_\mathcal{N}}{E_\nu^2} - \dfrac{2 T_\mathcal{N}}{E_\nu} + \dfrac{T_\mathcal{N}}{m_\mathcal{N}} - \dfrac{T_\mathcal{N}^2}{ m_\mathcal{N}E_\nu } - \dfrac{T_\mathcal{N}^2}{E_\nu^2} \right) + \dfrac{m_\chi^2}{2 E_\nu^2} \left(1 + \dfrac{3 E_\nu}{m_\mathcal{N}} - \dfrac{T_\mathcal{N}}{m_\mathcal{N}} + \dfrac{m_\chi^2}{m_\mathcal{N} T_\mathcal{N}}\right)\right] \tilde{S}^\mathcal{T}(\qtransfer^2)} \nonumber \\[4pt]
    &\hspace{-7em} \medmath{+ \left[\left(1 - \dfrac{T_\mathcal{N}}{E_\nu} -\dfrac{T_\mathcal{N}}{2 m_\mathcal{N}} + \dfrac{T_\mathcal{N}^2}{2 E_\nu^2} + \dfrac{T_\mathcal{N}^2}{2 m_\mathcal{N} E_\nu}\right) - \dfrac{m_\chi^2}{2 E_\nu^2} \left(1 + \dfrac{3 E_\nu}{2 m_\mathcal{N}} - \dfrac{T_\mathcal{N}}{2 m_\mathcal{N}} + \dfrac{m_\chi^2}{2 m_\mathcal{N} T_\mathcal{N}}\right)\right] \tilde{S}^\mathcal{L}(\qtransfer^2) \Bigg\rbrace\,} . \label{eq:cross-section-tensor-CEvNS}
\end{align}
The nuclear form factor accounts for nuclear-physics effects. We adopt the Klein-Nystrand parametrization\footnote{We assume that this form factor is the same for protons and neutrons. This is a good approximation for the typical momentum transfer involved at the COHERENT experiment.}, defined as
\begin{equation}\label{eq:form-factor-kn}
    F_W(\qtransfer^2) =  \dfrac{3\, j_1(\qtransfer R_A)}{\qtransfer R_A (1 + a_k^2 \qtransfer^2)} \, ,
\end{equation}
where $j_1(x) = \sin(x)/x^2 - \cos(x)/x$ is the spherical Bessel function of order one, $R_A = 1.23 \,A^{1/3}~\mathrm{fm}$ is the nuclear radius, $A$ is the mass number and $a_k = 0.7~\mathrm{fm}$ is the Yukawa potential range. The couplings $C_a$ are obtained following the procedure of \cite{Cirelli:2013ufw,DelNobile:2021wmp} and take the form
\begin{align}
\label{eq:couplingsCa}
    C_S^2 &\equiv g_S^2 \left( Z \sum_{q = u, d} \dfrac{m_p}{m_q} f_{T_q}^{(p)} + N \sum_{q = u,d} \dfrac{m_n}{m_q} f_{T_q}^{(n)} \right), \\[4pt]
    C_P^2 &\equiv g_P^2 \left[ Z \sum_{q = u, d} \dfrac{m_p}{m_q} \Delta_{q}^{(p)} + N \sum_{q = u,d} \dfrac{m_n}{m_q} \Delta_{q}^{(n)} \right] \left(1 - \sum_{q' = u, d} \dfrac{\overline{m}}{m_{q'}} \right), \\[4pt]
    C_V^2 &\equiv 3A g_V^2, \label{eq:couplingsCV}
\end{align}
where $Z$ is the number of protons, $N = A - Z$ is the number of neutrons, $m_p$ and $m_n$ are the proton and neutron masses, respectively, and $m_q$ are the quark $q$ masses. The constant $\overline{m}$ is defined as $1 / \overline{m} \equiv 1 / m_u + 1 / m_d + 1 / m_s$  (for further details, see Appendix~\ref{sec:hadronic-physics}). In the expressions for the axial-vector and tensor mediated cross sections~\footnote{Let us notice that the pseudoscalar interaction is also spin-dependent, however its CE$\nu$NS cross section is already suppressed due to its dependence on the nuclear recoil energy and for this reason we do not explicitly account for the nuclear-spin dependence in Eq.~\eqref{eq:cross-section-pseudoscalar-CEvNS}. This choice does not impact sizeably our results since the COHERENT bounds are dominated by the E$\nu$ES contribution.}, $J$ denotes the total angular momentum of the nucleus in the ground state, i.e., $5/2^+$ for I and $7/2^+$ for Cs. The spin structure functions $\tilde{S}^\kappa(\qtransfer^2)$, with $\kappa = \mathcal{L},\, \mathcal{T}$ to account for longitudinal and transverse multipoles calculated using the Shell Model are defined in Appendix~\ref{app:SpinSfunc}.

Let us also note that compared to conventional studies that rely upon phenomenological form factors to account for the finite nuclear size,  the use of different nuclear models may lead to differences in the expected event rates of up to 10\%~\cite{Hoferichter:2020osn,Kosmas:2021zve}. It is worth noting, however, that nuclear model-dependent uncertainties as large as $\sim 5\%$  also exist~\cite{Pandey:2023arh}. In our present work, the latter effects are effectively accounted for through the introduction of relevant nuisance parameters in the statistical analysis (see Sec.~\ref{sec:data-analysis}).

\section{Data analysis}\label{sec:data-analysis}
In this section, we provide a description of the procedure followed to analyze the most recent COHERENT, LZ and XENONnT datasets as well as the simulation of \eves~events at the DUNE-ND.

Given a neutrino flux with an energy endpoint $E_{\nu}^{\mathrm{max}}$, the maximum and minimum kinetic recoil energies that the atomic electrons in the detector can acquire, depend on the energy flux endpoint as follows
\begin{equation}\label{eq:recoil-energy-limits}
    T^{\prime \substack{\scriptscriptstyle \mathrm{min} \\ \scriptscriptstyle \mathrm{(max)}}}_{e}  = \dfrac{2m_{e}{(E_{\nu}^{\mathrm{max}})}^{2} - m_\chi^2\left(E_{\nu}^{\mathrm{max}} + m_{e}\right) {\substack{- \\ (+)}} E_{\nu}^{\mathrm{max}}\sqrt{4m_{e}^2 {(E_{\nu}^\mathrm{max})}^{2} - 4m_{e} m_\chi^2 \left(E_{\nu}^{\mathrm{max}} + m_{e}\right) + m_\chi^4}}{2m_{e}\left(2E_{\nu}^{\mathrm{max}} + m_{e}\right)}.
\end{equation}
Likewise, the minimum neutrino energy needed for the production of the SF is given, in terms of the true recoil energy of the electron, as
\begin{equation}\label{eq:minimum-neutrino-energy-limit}
    E_{\nu}^{\mathrm{min}} (T^\prime_e) = \dfrac{1}{2} \left(T^\prime_e + \sqrt{2 m_e T^\prime_e + (T^\prime_e)^2}\right) \left(1 + \dfrac{m_\chi^2}{2 m_e T^\prime_e}\right).
\end{equation}
In the case of nuclear recoil these expressions hold under the changes $m_e \rightarrow m_\mathcal{N}$ and $T_e \rightarrow T_{\mathcal{N}}$. Besides, in the limit $m_\chi \rightarrow 0$, the usual SM expressions are recovered.

\subsection{DUNE-ND}
For the case of DUNE-ND we  simulate the number of E$\nu$ES events  in the $E_e \theta_e^2$ space where $E_e = m_e + T_e$ and $\theta_e$ are the total reconstructed electron energy and scattering angle, respectively.  This way it becomes possible to remove a large part of the background which is mainly coming from charged-current quasielastic (CCQE) scattering and misidentified $\pi^0$, as explained in Ref.~\cite{MINERvA:2015nqi,Marshall:2019vdy}. From the kinematics of the process one has $1-\cos\theta_e = m_e(1-y) / E_e$ where $y$ denotes the inelasticity parameter, defined as $y\equiv T_e / E_\nu$ for SM E$\nu$ES. For the case of up-scattering into a SF it instead takes the form $y\equiv (T_e/ E_\nu) \left[1 + m_\chi^2 / (2 m_e T_e) \right]$. Then, the number of events as a function of $E_e \theta_e^2$ can be calculated as
\begin{equation} \label{eq:EvES_Differential_Event_Rate_EeTheta2}
\left[\dfrac{\mathrm{d}N}{\mathrm{d} E_e \theta_e^2}\right]_\lambda = t_\mathrm{run} \, N_e \, N_\mathrm{POT} \sum_{\ell}\int_{E_\nu^\mathrm{min}}^{E_\nu^\mathrm{max}}\mathrm{d}E_{\nu}\dfrac{\mathrm{d}\Phi_{\nu_\ell}(E_\nu)}{\mathrm{d}E_\nu}  \left[\dfrac{\mathrm{d}\sigma_{\nu_{\ell}}}{\mathrm{d} E_e \theta_e^2 }\right]_\lambda  \,,
\end{equation}
where the differential cross section reads
\begin{equation}
\dfrac{\mathrm{d}\sigma_{\nu_{\ell}}}{\mathrm{d} E_e \theta_e^2 } = \left.\dfrac{E_\nu}{2 m_e}\dfrac{\mathrm{d}\sigma_{\nu_{\ell}}}{\mathrm{d} T_e } \right\vert_{T_e = E_\nu \left(1 - \frac{E_e \theta_e^2}{2 m_e} - \Delta \right)}\, .
\end{equation}
Here, $\Delta=0$ when $\lambda=\mathrm{SM}$ and $\Delta=m_\chi^2/\left(2 m_e E_\nu \right)$ when $\lambda=\mathrm{SF}$.  In Eq.~\eqref{eq:EvES_Differential_Event_Rate_EeTheta2}, $\mathrm{d}\Phi_{\nu_\ell}(E_\nu) / \mathrm{d}E_\nu$ denotes  the Long Baseline Neutrino Facility (LBNF) neutrino flux for the different neutrino flavors $\nu_\ell = \{\nu_e, \nu_\mu, \bar{\nu}_e, \bar{\nu}_\mu\}$ and the various on-axis or off-axis locations~\cite{Fields:2017,DuneFluxes}.  We assume the DUNE-ND to be mobile up to 30 m off-axis~\cite{DUNE:2021tad}. Moreover, when simulating the neutrino spectra, for each neutrino flavor we consider both the nominal neutrino mode, namely the \emph{CP-optimized} beam and a  \emph{$\tau$-optimized} configuration~\cite{DUNE:2015lol,Machado:2020yxl}, the latter allowing for a shift of the neutrino spectra to higher energies\footnote{The CP-optimized configuration maximizes the neutrino flux at $E_\nu< 5$~GeV to optimize  CP-violating neutrino oscillation searches at the far detector. The $\tau$-optimized 
 configuration, instead, will produce a $\nu_\tau$  flux at the far detector that peaks at higher energies  compared to the standard one, e.g. for $5 > E_\nu > 10$~GeV, which in turn leads to an enhancement of   deep inelastic scattering events~\cite{DUNE:2020lwj, Ovchynnikov:2022rqj}.}.

Regarding the integration limits, the minimum neutrino energy to generate a detectable recoil is given by Eq.~\eqref{eq:minimum-neutrino-energy-limit}. The maximum neutrino energy $E_{\nu}^{\text{max}}$ is taken by the endpoint of the LBNF beam. Following~\cite{deGouvea:2019wav}, the signal spectra are evaluated assuming $50~\mathrm{MeV}  \leq E_e \leq 20~\mathrm{GeV}$  (as discussed in Ref.~\cite{Ballett:2019bgd}, concerning electron detection a threshold as low as 30~MeV is possible). 
Let us stress that a higher threshold of --- for example --- 600~MeV would modify our extracted sensitivities, at most, by a factor of 3 in the tensor case, whereas for the other interactions this change is milder. It should be kept in mind that the FNAL beam peaks at~2--3~GeV. Finally, the number of protons on target (POT) per year assuming a 120 GeV proton beam is taken to be $N_\mathrm{POT}=1.1 \times 20^{21}$~\cite{DUNE:2016hlj}, while $t_\text{run}$ is the exposure time and $N_e$ is the number of electron targets contained  in the LArTPC  of the DUNE-ND, for which we assume a fiducial volume of 67~tonnes. 

We evaluate the number of events by performing a Monte Carlo integration assuming an angular uncertainty of $\sigma_\theta=1^\mathrm{o}$, following the method described in Ref.~\cite{Melas:2023olz}. Before closing this discussion let us stress that we have verified that an uncertainty in the reconstruction of electron energy as large as $\sigma_E/E_e =10\%/\sqrt{E_e/\mathrm{GeV}}$ has negligible impact on the event rates, see e.g.~\cite{Mathur:2021trm,Chakraborty:2021apc,Melas:2023olz}.

To extract the sensitivity at the DUNE-ND, we rely on the analysis method of Ref.~\cite{Melas:2023olz}. A simultaneous  $\chi^2$ fit is performed taking into account on-axis (loc = 0 m) and off-axis (loc = 6, 12, 18, 24 and 30~m) spectra, using shape and normalization information, through the following $\chi^2$ function 
\begin{equation}
    \chi^2 = 2 \sum_{k = \nu, \,\bar{\nu}} \, \sum_{j= \mathrm{loc}} \, \sum_{i=1}^{20} \left[N^{\mathrm{th}}_{ijk} - N^{\mathrm{exp}}_{ijk}  + N^{\mathrm{exp}}_{ijk} \, \ln{\left(\dfrac{N^{\mathrm{exp}}_{ijk}}{N^{\mathrm{th}}_{ijk} } \right)} \right] + \left( \frac{\alpha_1}{\sigma_{\alpha_1}}\right)^2 + \left( \frac{\alpha_2}{\sigma_{\alpha_2}}\right)^2 \, .
\end{equation}
Here, the index $i$ runs over the reconstructed $E_e \theta_e^2$ bins for which we assume 20 bins in the range $[0, 10~m_e]$ as done in Ref.~\cite{Melas:2023olz}. The index $j$ runs over the different on-axis and off-axis locations, while the index  $k$ refers to the neutrino (3.5 years) and antineutrino mode (3.5~years). The expected theoretical number of events depends on the nuisance parameters $\alpha_1$ and $\alpha_2$ which account for the uncertainty on the neutrino flux ($\sigma_{\alpha_1}=5\%$) and background ($\sigma_{\alpha_2}=10\%$) normalizations, and reads $N_\mathrm{th} = (1+ \alpha_1) N_\mathrm{SM} + (1+\alpha_2) N_\mathrm{bkg}$.  On the other hand, $N_\mathrm{exp}$ accounts for the simulated data that DUNE-ND would detect including the new-physics SF events. Finally, the background events are taken to be the  sum of CCQE and misidentified $\pi^0$ events, i.e., $N_\text{bkg} = N_{\pi^0}^\text{missID} + N_\text{CCQE}$ (see Ref.~\cite{deGouvea:2019wav}). The latter events are weighted accordingly to match the exposure time at the various off-axis locations. We also note that although the shape of the background events is expected to vary for the different on-axis or off-axis locations, this effect is foreseen to be small.

\subsection{XENONnT and LUX-ZEPLIN}
Solar neutrinos induce low-energy electron recoils via E$\nu$ES events that are detectable at DM DD experiments.  We are interested in extracting limits on the up-scattering production of the SF using available data from the state-of-the-art experiments XENONnT and LZ. The signal can be obtained from the expression
\begin{equation}
   N^\mathrm{E\nu ES}_{i,k} = \mathcal{E}N_T~\int_{T_e^i}^{T_e^{i+1}} \mathrm{d}T_e ~ \epsilon_E(T_e)   ~\int_{T_e^{'{\mathrm{min}}}}^{ T_e^{'\mathrm{max}}} \mathrm{d}T_e'~ R(T_e,T_e')~
   \sum_{k=pp,\,^{7}\mathrm{Be}}\int_{E_{\nu}^{\mathrm{min}}}^{E_{\nu,k}^{\mathrm{max}}} \mathrm{d}E_\nu ~\sum_{\ell} ~\dfrac{\mathrm{d}\Phi_{\nu_{\ell}}^{k}}{\mathrm{d} E_\nu}~ \dfrac{\mathrm{d}\sigma_{\nu_{\ell} \mathcal{A}}}{\mathrm{d}T_e'}\, ,
\label{eq:number-events-lz-xenon}
\end{equation}
where $T_e$ and $T_e'$ are the reconstructed and true electron recoil energies, respectively, and $E_\nu$ is the neutrino energy. Here, $\mathcal{E}$ represents the exposure given in units of $\mathrm{tonnes \times years}$, while $N_T$ corresponds to the number of target nuclei (LXe in the case of XENONnT and LZ) per $\mathrm{tonne}$.  For XENONnT (LZ) the exposure time is $\mathcal{E} = 1.16~ (0.90)~\mathrm{tonne \times year}$ (see Refs.~\cite{XENON:2022ltv,LZ:2022lsv}). The number of electron targets depends on the value of the energy-dependent quantity, $Z_\mathrm{eff}^\mathcal{A}$, which is taken into account in the cross section. 

The lower integration limit for the neutrino energy is given by Eq.~\eqref{eq:minimum-neutrino-energy-limit}. On the other hand, the upper limit $E_{\nu,k}^\mathrm{max}$ in this case corresponds to the endpoints of the solar neutrino energy spectra and in particular for $k = pp,\, ^{7}\mathrm{Be}(0.861)$,  the two relevant flux components given the current detector thresholds. Likewise, the true recoil energy limits are given by Eq.~\eqref{eq:recoil-energy-limits}.

The fluxes $\mathrm{d} \Phi_{\nu_{\ell}}^{k} / \mathrm{d} E_\nu$ are given by 
\begin{equation}
\Phi_{\nu_{e}}^{k}
=
\Phi_{\nu_{e}}^{k\,\odot} P_{ee},
\quad
\Phi_{\nu_{\mu}}^{k}
=
\Phi_{\nu_{e}}^{k\,\odot} \left( 1 - P_{ee} \right) \cos^2\theta_{23},
\quad
\Phi_{\nu_{\tau}}^{k}
=
\Phi_{\nu_{e}}^{k\,\odot} \left( 1 - P_{ee} \right) \sin^2\theta_{23},
\label{eq:solflux}
\end{equation}
where $\Phi_{\nu_{e}}^{k\,\odot}$ denote the unoscillated fluxes of $\nu_{e}$ produced
in the interior of the Sun~\cite{bahcall_web,Bahcall:1987jc,Bahcall:1994cf,Bahcall:1996qv}, while the index $k$  corresponds to the various production mechanisms.  As anticipated, for XENONnT and LZ  only  $k =pp, \, ^{7}\text{Be}$ are relevant, whose corresponding flux normalizations are taken from Ref.~\cite{Baxter:2021pqo}.
 In Eq.~(\ref{eq:solflux}), $P_{ee}$ represents the $\nu_{e}$-survival probability at the terrestrial detector accounting for neutrino oscillations which we take from Ref.~\cite{AristizabalSierra:2017joc}, while $\theta_{23}$ stands for the atmospheric mixing angle, see e.g.~\cite{deSalas:2020pgw}.
 
Finally, the quantities $R(T_e,T_e')$ and $\epsilon_E(T_e)$ denote the detector resolution and efficiency, respectively. For the case of XENONnT we employ the resolution function from Ref.~\cite{XENON:2020rca} and the efficiency from Ref.~\cite{XENON:2022ltv}. Similarly, for the case of LZ the resolution is taken from Refs.~\cite{AtzoriCorona:2022jeb,A:2022acy} and the efficiency from Ref.~\cite{LZ:2023poo}.

Next, let us analyze the most recent data reported by LZ and XENONnT collaborations. Regarding the LZ experiment, our statistical analysis is based on the following Poissonian $\chi^2$ function~\cite{DeRomeri:2024dbv} 
\begin{equation}
    \chi^2 = 2\left[\sum_{i=1}^{51} N^{\mathrm{th}}_{i} - N^{\mathrm{exp}}_{i} + N^{\mathrm{exp}}_{i} \ln\left(\dfrac{N^{\mathrm{exp}}_{i}}{N^{\mathrm{th}}_{i}}\right)\right] + \sum_k\left(\dfrac{\alpha_k}{\sigma_{\alpha_k}}\right)^2 + \sum_k\left(\dfrac{\beta_k}{\sigma_{\beta_k}}\right)^2 \, ,
    \label{eq:chi2_LZ}
\end{equation}
where $N^{\mathrm{exp}}_{i}$ is the data reported in Ref.~\cite{LZ:2022lsv} in the $i$-th  recoil-energy bin, $\alpha_k$ and $\beta_k$ are nuisance parameters of the backgrounds and neutrino fluxes, respectively, and $\sigma_{\alpha_k}$, $\sigma_{\beta_k}$ are their associated uncertainties. Such uncertainties are given in Ref.~\cite{LZ:2022ysc} for $\sigma_{\alpha_k}$ and in Ref.~\cite{Baxter:2021pqo} for $\sigma_{\beta_k}$. The predicted number of events, $N^{\mathrm{th}}_{i}$, for the energy bin $i$ is given by
\begin{equation}
    N^{\mathrm{th}}_{i} = \sum_{k=pp,\,^{7}\mathrm{Be}} (1 + \beta_k) N^\mathrm{E\nu ES}_{i,k} + \sum_k (1 + \alpha_k) N_{i,k}^{\mathrm{bkg}},
\end{equation}
where $N^\mathrm{E\nu ES}_{i,k}$ is given by Eq.~\eqref{eq:number-events-lz-xenon} for each flux $k$ and $N_{i,k}^{\mathrm{bkg}}$ is every background described in Table~VI of Ref.~\cite{LZ:2022ysc}. At this point we should stress that although we are varying all the background components individually according to Eq.~(\ref{eq:chi2_LZ}) as done in Ref.~\cite{DeRomeri:2024dbv}, our present results would differ by less than 5\% if we followed the method of Ref.~\cite{A:2022acy}
where all the backgrounds were summed together and a single nuisance parameter was used. We however, stick to the more complete method of Ref.~\cite{DeRomeri:2024dbv} despite the computational cost. For the case of XENONnT experiment, we instead rely on the Gaussian $\chi^2$ function
\begin{equation}
    \chi^2= \sum_{i=1}^{30} \left( \dfrac{N^{\mathrm{th}}_{i} - N^{\mathrm{exp}}_{i}}{\sigma_i} \right)^2 +\sum_k\left(\dfrac{\beta_k}{\sigma_{\beta_k}}\right)^2 \, ,
\end{equation}
where $N^{\mathrm{exp}}_{i}$ and $\sigma_i$ denote the experimental data and the corresponding uncertainties, respectively, taken from Ref.~\cite{XENON:2022ltv}. The nuisance parameters $\beta_k$ and their uncertainties $\sigma_{\beta_k}$ are the same as those of Eq.~\eqref{eq:chi2_LZ}. Moreover,
\begin{equation}
    N^{\mathrm{th}}_{i} = \sum_k (1 + \beta_k) N^\mathrm{E\nu ES}_{i,k} + B_i^0 \, .
\end{equation}
Here, $B_i^0$ represents the modeled background reported in~\cite{XENON:2022ltv} from which the SM E$\nu$ES contribution has been removed, and $N^\mathrm{E\nu ES}_{i,k}$ is obtained with Eq.~\eqref{eq:number-events-lz-xenon}. While in principle the more sophisticated method of Ref.~\cite{DeRomeri:2024dbv} could be adopted like we did for the case of LZ, here we choose the other one for simplicity. Indeed, the  two methods give anyway identical results, compare, e.g., the results of Refs.~\cite{A:2022acy,DeRomeri:2024dbv}. 

\subsection{COHERENT}
In the case of the COHERENT experiment we analyze the most recent data released by the COHERENT collaboration from the CsI measurement~\cite{COHERENT:2021xmm}\footnote{The COHERENT collaboration has also released results using a LAr target~\cite{COHERENT:2020iec}. However, we do not analyze this data set here as it currently has less statistics and hence leads to less stringent bounds than the CsI-2021 data set (see e.g. Refs.~\cite{DeRomeri:2022twg,Candela:2023rvt}).}. We perform a comprehensive analysis exploiting both the energy and time data from the COHERENT-CsI-2021 measurement following the fitting procedure described in Refs.~\cite{DeRomeri:2022twg,Candela:2023rvt}.

First, we compute the number of SF or SM events expected at the CsI detector, of mass $m_{\mathrm{det}} = 14.6~\mathrm{kg}$, and located at a distance $L = 19.3~\mathrm{m}$ from the neutrino source, i.e., the Spallation Neutron Source (SNS). The expected number of events on a nuclear target, $\mathcal{N}$, per neutrino flavor, $\nu_\ell$, in each nuclear recoil energy bin, $i$, is given by
\begin{align}
\label{eq:Nevents_SF_CEvNS}
N_{i, \nu_\ell}^{\mathrm{CE\nu NS}}(\mathcal{N})
= \nonumber
& \, N_\mathrm{target}
\int_{T_\mathcal{N}^i}^{T_\mathcal{N}^{i+1}}
\hspace{-0.3cm}
\d T_\mathcal{N}\,
\epsilon_E(T_\mathcal{N})
\int_{T_\mathcal{N}^{'{\mathrm{min}}}}^{ T_\mathcal{N}^{'\mathrm{max}}}
\hspace{-0.3cm}
\d T_\mathcal{N}^\prime
\,
R(T_\mathcal{N}, T_\mathcal{N}^\prime)  \\
& \times \int_{E_\nu^{\mathrm{min}}(T_\mathcal{N}^\prime)}^{E_\nu^{\mathrm{max}}}
\d E_\nu \,
\dfrac{\d \Phi_{\nu_\ell}}{\d E_\nu}(E_\nu)
\dfrac{\d \sigma_{\nu_\ell \mathcal{N}}}{\d T_\mathcal{N}^\prime}(E_\nu, T_\mathcal{N}^\prime) \, 
,
\end{align}
where $N_\mathrm{target} = N_{\mathrm{A}} m_{\mathrm{det}} / M_{\mathrm{\mathrm{target}}}$ is the number of target nuclei in the detector. The integration limits for $T_\mathcal{N}^\prime$ and $E_\nu$ are given by Eqs.~\eqref{eq:recoil-energy-limits} and \eqref{eq:minimum-neutrino-energy-limit}. Moreover, $\epsilon_E(T_{\mathcal{N}})$ is the energy-dependent detector efficiency, $R(T_\mathcal{N}, T_\mathcal{N}^\prime)$ is the energy resolution function and $\mathrm{d}\Phi_{\nu_\ell} / \mathrm{d}E_\nu$ denote the prompt ($\nu_\mu$) and delayed ($\nu_e$ and $\overline{\nu}_\mu$) neutrino fluxes produced at the SNS, for which we employ the well-known Michel spectra. The nuclear-recoil spectrum given in Eq.~(\ref{eq:Nevents_SF_CEvNS}) is then converted to an electron-equivalent spectrum using the scintillation response function provided in Ref.~\cite{COHERENT:2021xmm} via the quenching factor (QF) measurements reported in~\cite{COHERENT:2021pcd}. In order to be able to directly compare our results with the experimental data, in a final step we need to eventually convert the electron-recoil spectrum into a photoelectron (PE) spectrum using the light yield information, as explained in Ref.~\cite{COHERENT:2021xmm}.

Since the CsI detector cannot distinguish between a nuclear or an electron recoil, we also include in the analysis the corresponding expected number of \eves~events on an atomic nucleus $\mathcal{A}$, estimated as
\begin{align}
\label{eq:Nevents_SF_ES}
N_{i, \nu_\ell}^{\mathrm{E\nu ES}}(\mathcal{A})
= \nonumber
& \, N_\mathrm{target}
\int_{T_e^i}^{T_e^{i+1}}
\hspace{-0.3cm}
\d T_e\,
\epsilon_E(T_e)
\int_{T_e^{'{\mathrm{min}}}}^{ T_e^{'\mathrm{max}}}
\hspace{-0.3cm}
\d T_e^\prime
\,
R(T_e, T_e^\prime)  \\
& \times \int_{E_\nu^{\mathrm{min}}(T_e^\prime)}^{E_\nu^{\mathrm{max}}}
\d E_\nu \,
\dfrac{\d \Phi_{\nu_\ell}}{\d E_\nu}(E_\nu)
\dfrac{\d \sigma_{\nu_\ell \mathcal{A}}}{\d T_e^\prime}(E_\nu, T_e^\prime) \, .
\end{align}
Similar to the \cevns~case, the computed \eves~spectrum is eventually converted into a PE-spectrum to be directly confronted to the experimental data.

In addition, we include timing information in the analysis by distributing the predicted number of events $N_{i, \nu_\ell}^{\mathrm{CE\nu NS}}(\mathcal{N})$ in each time bin $j$. We adopt the time distributions $\mathcal{P}^{\nu_\ell}_T(t_{\mathrm{rec}})$ of $\nu_\ell = \nu_\mu,\, \overline{\nu}_\mu,\, \nu_e$ given in~\cite{Picciau:2022xzi,COHERENT:2021xmm}, normalized to 6 $\mathrm{\mu s}$. Thus, the expected number of events per each observed nuclear recoil energy bin $i$ and time bin $j$ is given by
\begin{equation}
\label{eq:Nevents_SF_ij_CEvNS}
 N_{ij}^{\mathrm{CE\nu NS}}(\mathcal{N}) = \sum_{\nu_\ell =\nu_{e}, \nu_{\mu}, \bar{\nu}_{\mu}}\int_{t_{\mathrm{rec}}^{j}}^{t_{\mathrm{rec}}^{j+1}} \d t_{\mathrm{rec}} \, \mathcal{P}^{\nu_\ell}_T(t_{\mathrm{rec}}, \alpha_6)\,\epsilon_T(t_{\mathrm{rec}}) N_{i, \nu_{\ell}}^{\mathrm{CE\nu NS}}(\mathcal{N}),
\end{equation}
where $t_\mathrm{rec}$ is the reconstructed time and $\epsilon_T(t_{\mathrm{rec})}$ is the time-dependent efficiency. We have included an additional nuisance parameter on the beam timing, $\alpha_6$, to account for the uncertainty on the beam arrival time (further details are provided at the end of this discussion). For the \eves~rate, the timing information is also included using Eq.~\eqref{eq:Nevents_SF_ij_CEvNS} but by exchanging the predicted number of CE$\nu$NS events with the predicted number of \eves~events given in Eq.~\eqref{eq:Nevents_SF_ES}. Finally, the SF and SM contributions must be summed, and a weighted sum for the Cs and I nuclei must be performed to obtain the total number of expected events ($ N_{ij}^{\mathrm{CE\nu NS}}$ and $N_{ij}^{\mathrm{E \nu ES}}$ for CE$\nu$NS and \eves, respectively).

For the statistical analysis of the COHERENT-CsI-2021 data we employ the following Poissonian least-squares function \cite{DeRomeri:2022twg}
\begin{equation}\label{eq:chi2CsI}
	\chi^2_{\mathrm{CsI}}\Big|_{\mathrm{CE\nu NS} + \mathrm{E \nu ES}}
	 =
	2
	\sum_{i=1}^{9}
	\sum_{j=1}^{11}
	\left[ N^\mathrm{th}_{ij}  -  N_{ij}^{\mathrm{exp}} 
	 +  N_{ij}^{\mathrm{exp}} \ln\left(\frac{N_{ij}^{\mathrm{exp}}}{ N^\mathrm{th}_{ij}} \right)\right]\\
	+ \sum_{k=0}^{5}
	\left(
	\dfrac{ \alpha_{k} }{ \sigma_{k} }
	\right)^2  
	,
\end{equation}
where $N_{ij}^{\mathrm{exp}}$ is the binned coincidence data given in \cite{COHERENT:2021xmm} and $N^\mathrm{th}_{ij}$ is the predicted number of events, given by
\begin{align}\nonumber
N^\mathrm{th}_{ij} = &(1 + \alpha_{0} +\alpha_{5})  N_{ij}^{\mathrm{CE\nu NS}} (\alpha_{4}, \alpha_{6}, \alpha_{7})  + (1 + \alpha_{0} )  N_{ij}^{\mathrm{E \nu E S}} (\alpha_{6}, \alpha_{7})\\[4pt]
 &+ (1 + \alpha_{1}) N_{ij}^\mathrm{BRN}(\alpha_{6}) + (1 + \alpha_{2}) N_{ij}^\mathrm{NIN}(\alpha_{6})
 + (1 + \alpha_{3}) N_{ij}^\mathrm{SSB}  \,.
	\label{eq:Nth_CsI_chi2}
\end{align}
Here, the number of events for each background are $N_{ij}^\mathrm{BRN}$, $N_{ij}^\mathrm{NIN}$ and $N_{ij}^\mathrm{SSB}$, which stand for beam related neutrons (BRN), neutrino induced neutrons (NIN) and steady state backgrounds (SSB), respectively, all taken from~\cite{COHERENT:2021xmm}. Equations~\eqref{eq:chi2CsI} and \eqref{eq:Nth_CsI_chi2} involve several nuisance parameters ($\alpha_i$) together with their associated uncertainties ($\sigma_i$). In particular, $\sigma_{0} = 11\%$  encodes efficiency and flux uncertainties; $\sigma_{1} = 25\%$, $\sigma_{2} = 35\%$ and $\sigma_{3} = 2.1\%$ are related to BRN, NIN and SSB backgrounds, respectively. On the other hand, $\sigma_{5} = 3.8\%$ is associated to the QF. Notice that the number of events $N_{ij}^{\mathrm{CE\nu NS}}$ and $N_{ij}^{\mathrm{ES}}$ further include: $\alpha_{4}$, that enters the nuclear form factor and thus affects only the CE$\nu$NS number of events~\footnote{This is done by introducing the nuisance parameter $\alpha_4$ in the nuclear radius entering Eq.~\eqref{eq:form-factor-kn} such that $R_A = 1.23 \, A^{1/3} (1+\alpha_4)$.}, with  $\sigma_{4} = 5\%$; $\alpha_{6}$, that accommodates the uncertainty in beam timing with no prior assigned; and $\alpha_{7}$, that allows for deviations of the uncertainty in the \cevns ~efficiency. We refer the reader to Ref.~\cite{DeRomeri:2022twg} for further details regarding the statistical analysis. 

\section{Results and discussion}
\label{sec:results}
In this section, we summarize the results of our analyses, starting with the prospects for DUNE-ND and then comparing the constraints inferred using current data from the COHERENT, XENONnT and LZ experiments.

\subsection{DUNE-ND sensitivities}
\begin{figure}
    \centering
    \begin{subfigure}{0.49\textwidth}
        \includegraphics[width=\textwidth]{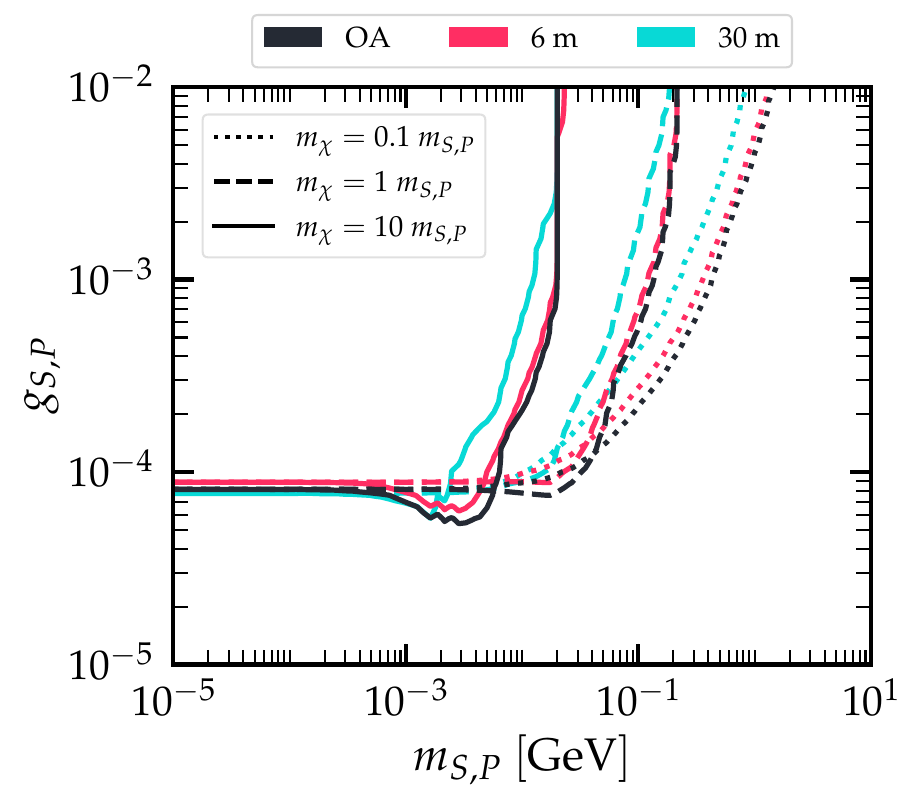}
    \end{subfigure}
    \hfill
    \begin{subfigure}{0.49\textwidth}
        \includegraphics[width=\textwidth]{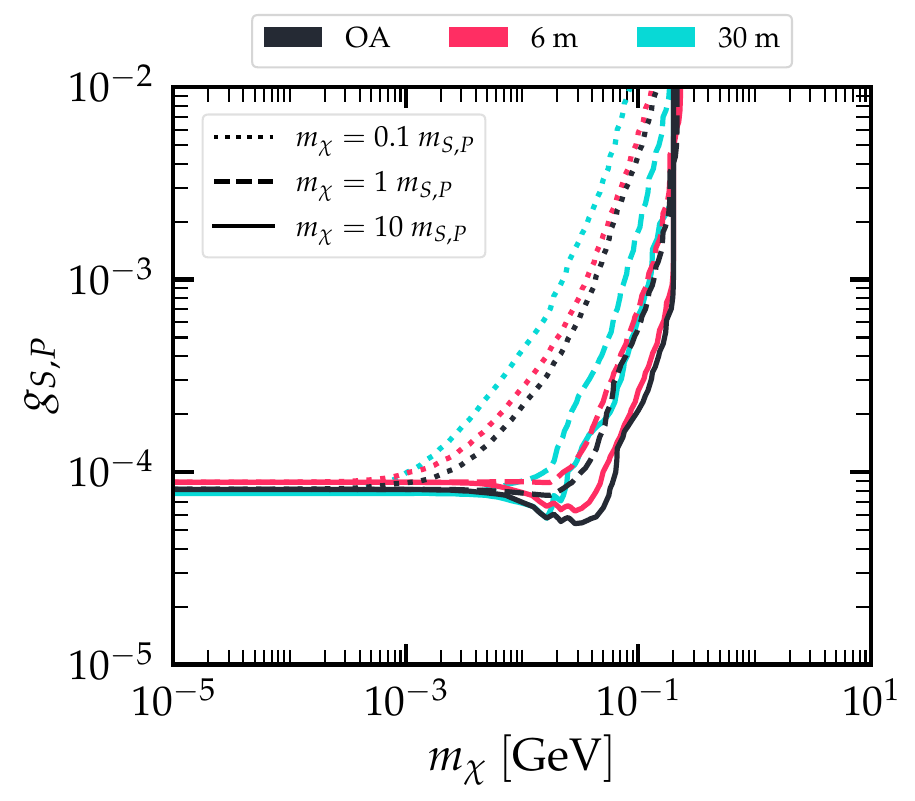}
    \end{subfigure}
    
    \begin{subfigure}{0.49\textwidth}
        \includegraphics[width=\textwidth]{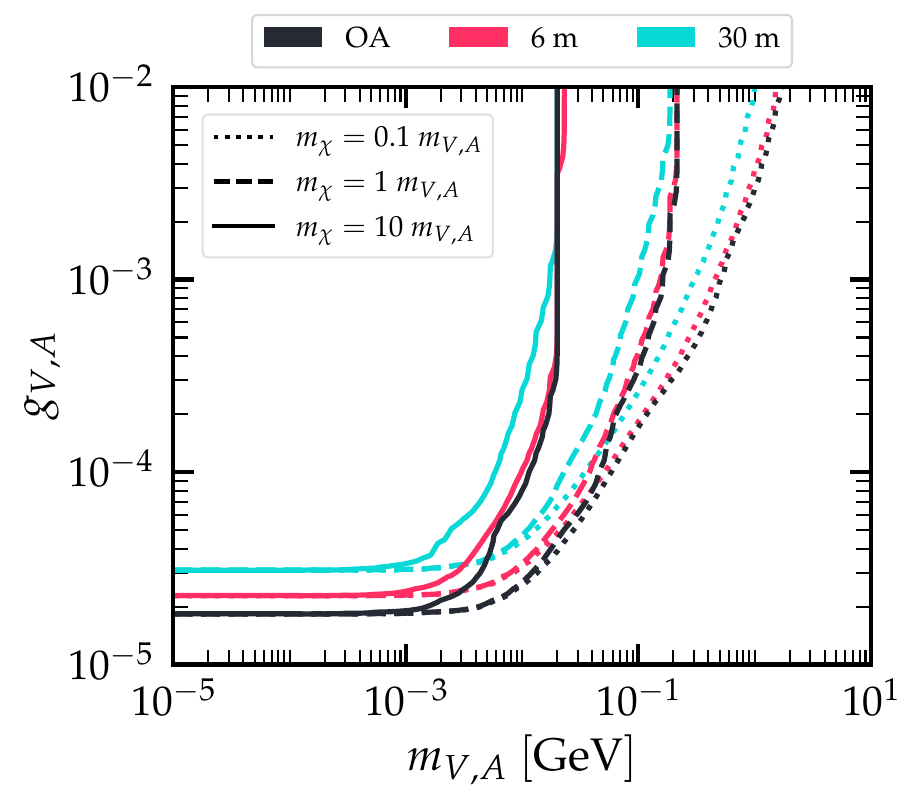}
    \end{subfigure}
    \hfill
    \begin{subfigure}{0.49\textwidth}
        \includegraphics[width=\textwidth]{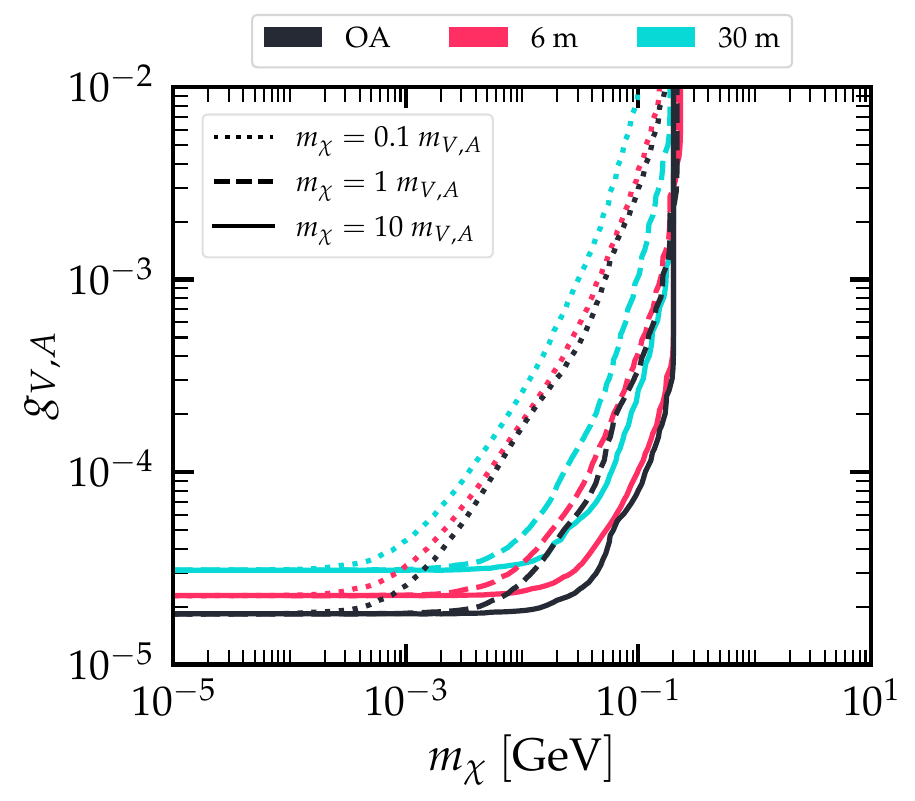}
    \end{subfigure}
    
    \begin{subfigure}{0.49\textwidth}
        \includegraphics[width=\textwidth]{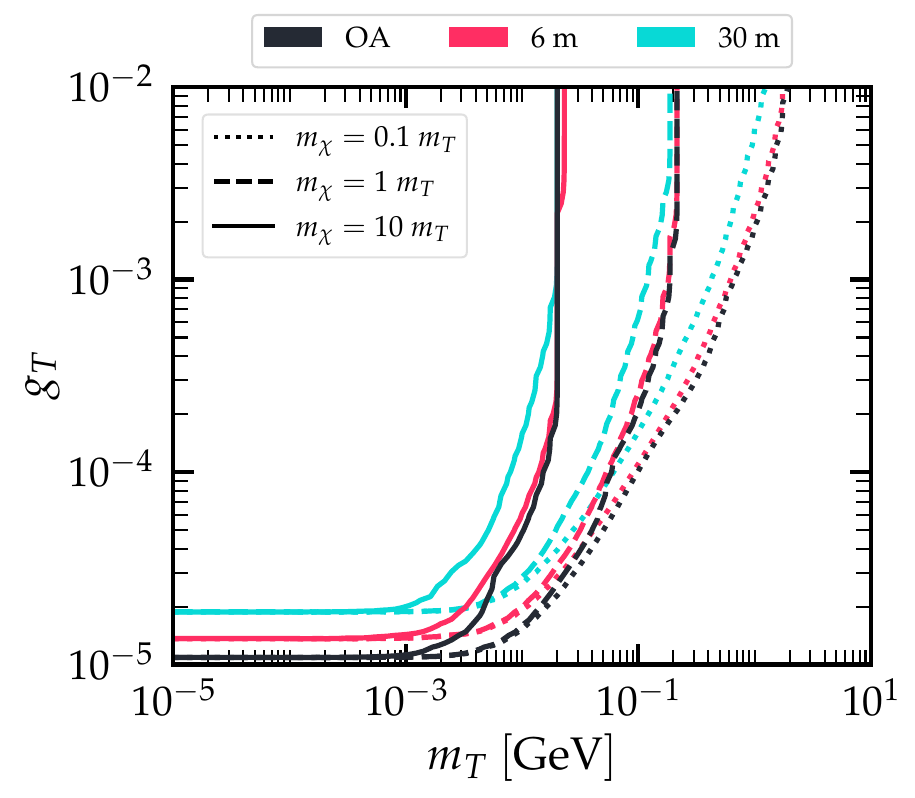}
    \end{subfigure}
    \hfill
    \begin{subfigure}{0.49\textwidth}
        \includegraphics[width=\textwidth]{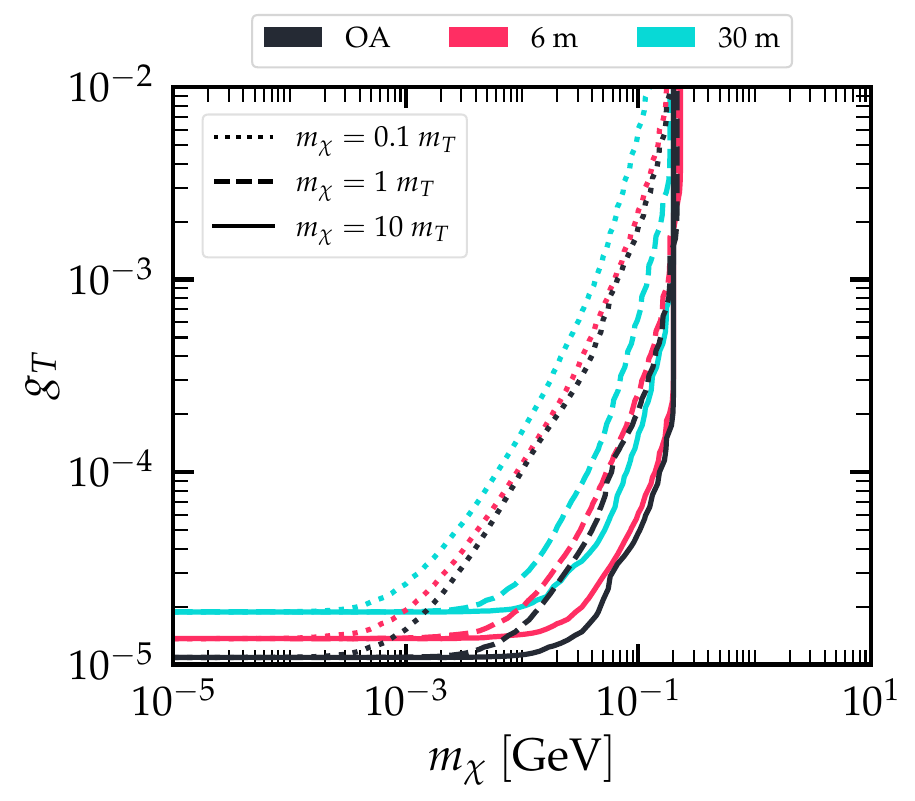}
    \end{subfigure}
    \caption{90\% C.L. projected sensitivity of DUNE-ND on the SF parameter space, after 7 years (3.5 in neutrino mode and 3.5 in antineutrino mode) of data taking time.  Individual sensitivities corresponding to some on-axis (OA) and off-axis (6 and 30~m) locations are shown. The results are presented for the various possible interactions: scalar/pseudoscalar, vector/axial-vector and tensor. Dotted, dashed and plain curves account for the three benchmark scenarios: $m_\chi= \{0.1, \,1, \,10\} \times \, m_a$. Left (Right) panels depict the sensitivity regions projected on the mediator (SF) mass.}
    \label{fig:dune-locations}
\end{figure}

In Fig.~\ref{fig:dune-locations} we present the 90\% C.L. sensitivities on the up-scattering production of a SF at the DUNE-ND for various  on-axis and off-axis locations. We assume 7~years of total running time (3.5~years in neutrino mode and 3.5~years in antineutrino mode). The calculation takes into account the actual exposure time in each location i.e., 50\% of running time is assumed to be on-axis and 10\% to each of the off-axis locations (6,~12,~18,~24~and~30)~m. For each location the corresponding sensitivity is obtained (for a combined analysis, see below). To avoid overcrowding the plots, here we depict only the cases correspondent to on-axis, 6~m and 30~m. The results are presented for the different interaction channels $a \equiv \left\{S,\, P,\, V,\, A,\, T\right\}$. For illustration purposes three benchmark choices $m_\chi= \{0.1, \,1, \,10\} \times \, m_a$ are depicted. The left panels are plotted by projecting the contours onto the  relevant coupling [see Eqs.~\eqref{eq:cross-section-scalar-ES} through \eqref{eq:cross-section-tensor-ES}] and mediator mass $m_a$, while the right panels are plotted with respect to the SF mass $m_\chi$. For small masses of the mediator, $m_a$, or of the SF, $m_\chi$, the constraints are reaching a saturation point, regardless on whether the contours are projected with respect to $m_a$ or $m_\chi$. On the other hand, for relatively large $m_a$ or $m_\chi$, the contours behave differently. Specifically,  the sensitivity loss occurs at different mediator masses $m_a$, depending on the relation between $m_a$ and $m_\chi$ (see left panels). However, when the results are projected with respect to $m_\chi$, the sensitivity loss occurs always at the same SF mass, due to the kinematics of the process\footnote{By imposing the conservation of energy and momentum, we find that the mass of the SF is constrained from above as $m_\chi \leq \sqrt{(M (M + 2 E_{\nu,\mathrm{max}})} - M$, where $M$ is the mass of the fixed target ($m_e$ or $m_\mathcal{N}$) and $E_{\nu,\mathrm{max}}$ is the maximum value of the incoming neutrino energy.}. 

\begin{figure}
    \centering
    \begin{subfigure}{0.49\textwidth}
        \includegraphics[width=\textwidth]{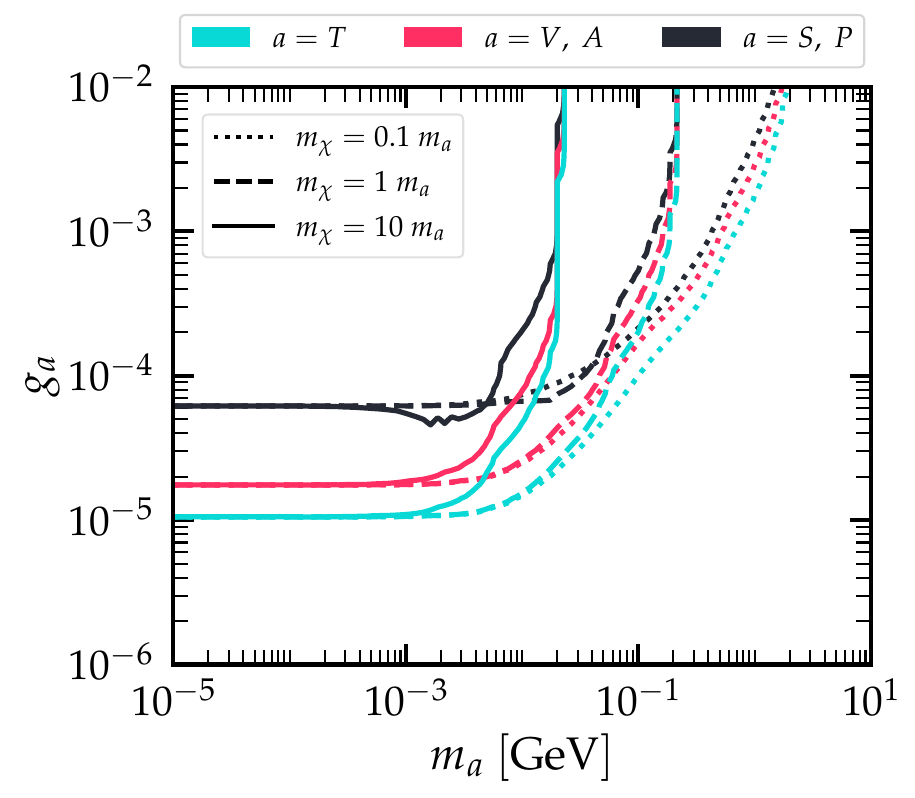}
    \end{subfigure}
    \hfill
    \begin{subfigure}{0.49\textwidth}
        \includegraphics[width=\textwidth]{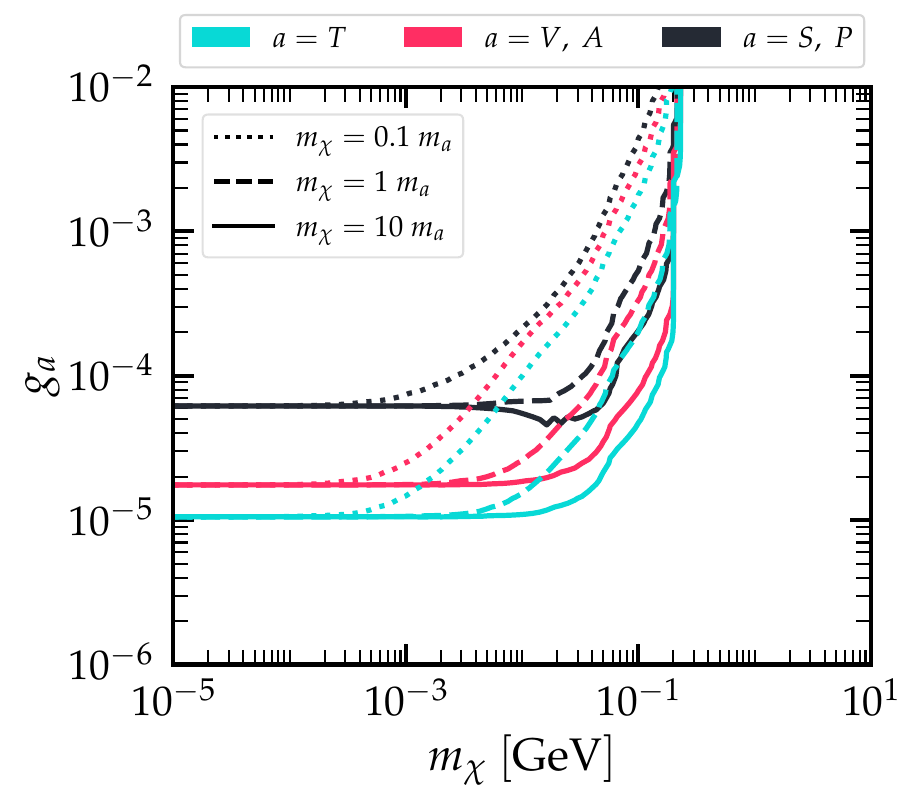}
    \end{subfigure}
    \caption{Comparison of the expected DUNE-ND sensitivities at 90\% C.L. after 7 years of data-taking time. The results are combined over all possible locations, accounting for the different exposure times at each location. The exclusion regions are presented in terms of the mediator mass $m_a$ (left) and the SF mass $m_\chi$ (right) for the three benchmark scenarios (see the main text for details).}
    \label{fig:DUNE_interactions_comparison}
\end{figure}

As can be seen in Fig.~\ref{fig:dune-locations}, the on-axis location yields the most stringent results in all cases, while the remaining off-axis locations yield more or less similar results which become successively less stringent as we move to further off-axis locations. This  was expected, since more neutrinos per unit time reach the DUNE-ND when the latter is placed on-axis than off-axis. The advantage of going off-axis is to reduce backgrounds, although at the price of also reducing the signal. For completeness, although not shown here, we have also examined the impact of the different neutrino flux shape distributions expected at the various locations of DUNE-ND. We did so by normalizing them appropriately by hand such that they all have the same flux normalizations. In this case we concluded that for all DUNE-ND locations the extracted constraints become identical, hence rendering the neutrino flux shape not relevant at the scope of this analysis.

Moreover, we note that the results for scalar/pseudoscalar or vector/axial-vector are presented with the same contour curves. This is due to  the following reasons. The cross sections for scalar and pseudoscalar cases have equivalent forms with the only difference being that the scalar is proportional to  $1+T_e/(2 m_e)$ while the pseudoscalar one is proportional to  $T_e/(2 m_e)$. However, for the GeV recoil energies involved at DUNE-ND it holds that $T_e/(2 m_e) \gg 1$, rendering the two cross sections practically equivalent. Note that this is not the case for the analysis of DM DD experiments where keV recoil energies are involved. The expressions of vector and axial-vector cross sections differ by relative signs in terms that contribute negligibly to the event rates. In particular the term $m_e T_e/(2 E_\nu^2)$ appearing in the first parenthesis of Eqs.~\eqref{eq:cross-section-vector-ES} and~\eqref{eq:cross-section-axialvector-ES} is $\ll 1$, while in the second parenthesis both $E_\nu/m_e$ and $T_e/m_e$ are $\gg 1$.

We are now interested in estimating the combined DUNE-ND sensitivities after 7 years of running time. Our result is depicted in Fig.~\ref{fig:DUNE_interactions_comparison} where we also compare the sensitivity reach of the different interaction channels. As previously, we consider the three benchmark choices $m_\chi= \{0.1, \,1, \,10\} \times \, m_a$, while left and right panels illustrate the contours with respect to  $m_a$ and $m_\chi$, respectively, with the same behavior as discussed in Fig.~\ref{fig:dune-locations}. From the plots it can be deduced that the scalar/pseudoscalar (tensor) interactions will be constrained with the least (most) sensitivity. Indeed, at the leading order, the scalar/pseudoscalar cross section is suppressed by a factor $T_e \left(m_\chi^2  + 2 m_e T_e\right)/\left(4 E_\nu^2 m_e\right)$ compared to the vector/axial-vector and tensor cross sections. On the other hand, the tensor cross section is by a factor 8 larger compared to the vector/axial-vector one.  Finally, let us comment on the marginal improvement of the combined analysis presented in Fig.~\ref{fig:DUNE_interactions_comparison}, compared to Fig.~\ref{fig:dune-locations}. As already explained, the best sensitivity is obtained when the DUNE-ND is on-axis and the neutrino flux is maximal. Indeed, the combined analysis is mainly driven by the on-axis events with a small improvement  coming from the contributions corresponding to off-axis events. 

\subsection{Comparing with COHERENT, XENONnT and LZ}
We now proceed to explore the complementarity of a future DUNE-ND measurement with existing data from \cevns~and DM DD experiments. In particular, we analyze the recent  \cevns~data from the COHERENT CsI-2021 measurement as well as the \eves~data from the DM DD experiments XENONnT and LZ. To this purpose we reproduce our DUNE-ND sensitivities discussed previously in Fig.~\ref{fig:DUNE_interactions_comparison} and we superimpose the corresponding exclusion regions in the light of existing COHERENT, XENONnT and LZ data. 

\begin{figure}[!htb]
    \centering
    \begin{subfigure}{0.49\textwidth}
        \includegraphics[width=\textwidth]{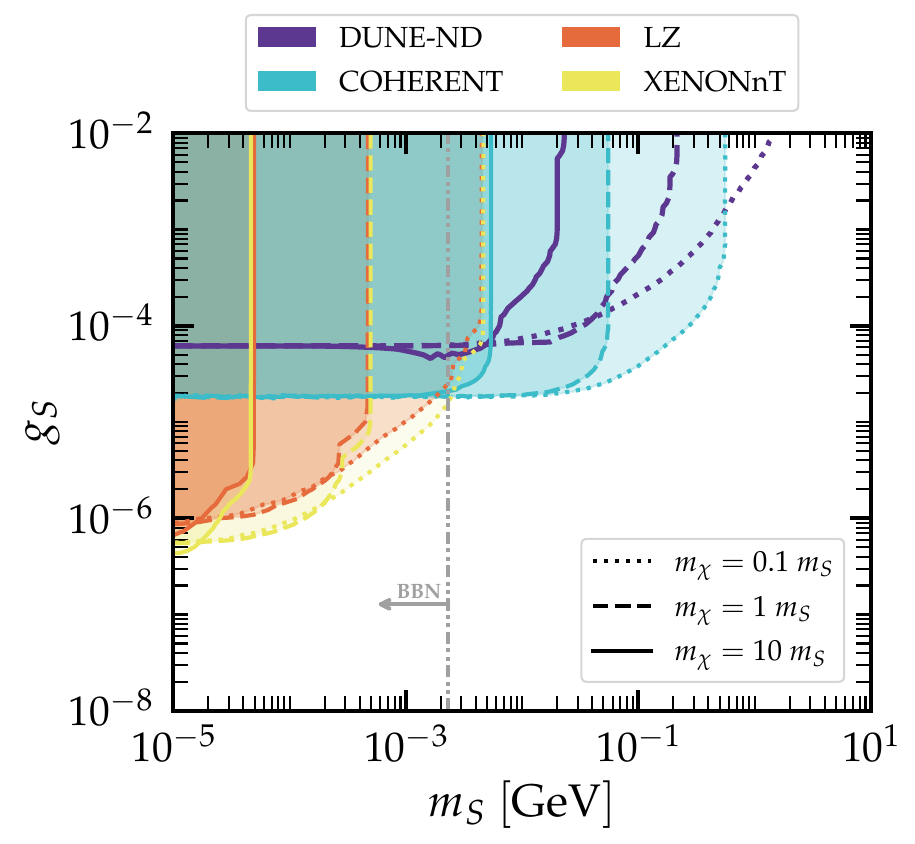}
    \end{subfigure}
    \hfill
    \begin{subfigure}{0.49\textwidth}
        \includegraphics[width=\textwidth]{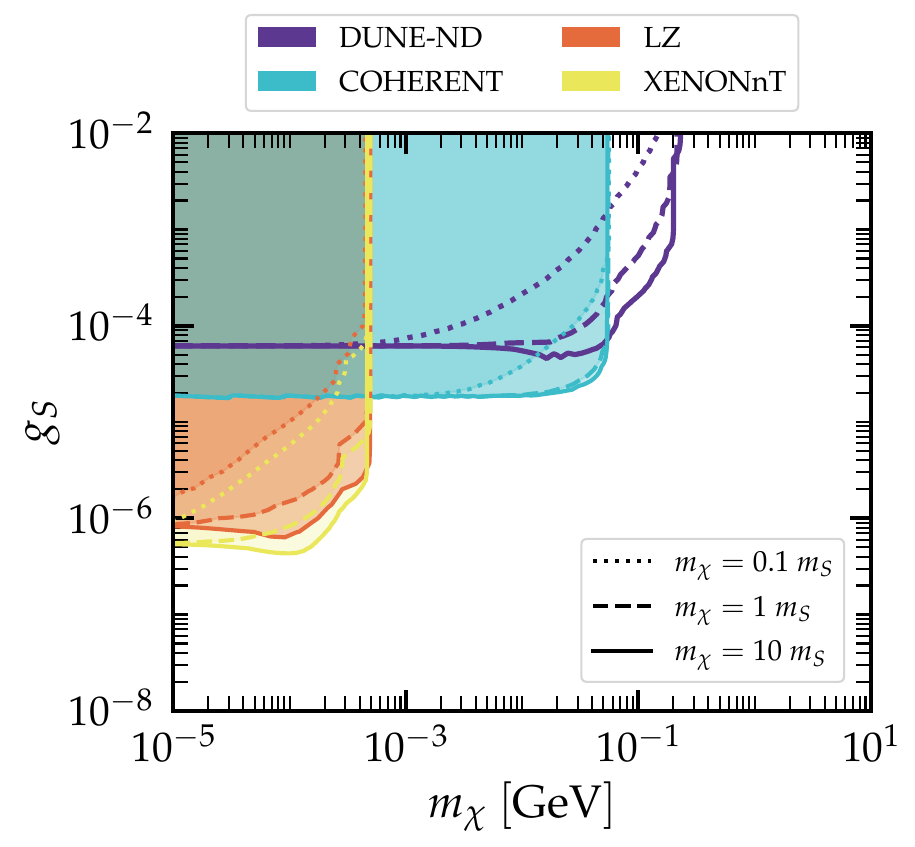}
    \end{subfigure}
    \caption{90\% C.L. exclusion regions for the case of scalar-mediated SF production. A comparison is given between the constraints coming from the analysis of existing data by XENONnT, LZ and COHERENT with the projected sensitivity from future measurements at the DUNE-ND. Left (Right) panels show the exclusion regions projecting on the mediator (SF) mass, for the three benchmark scenarios $m_\chi= \{0.1, \,1, \,10\} \times \, m_a$ considered.}
    \label{fig:all-experiments-scalar}
\end{figure}

\begin{figure}
    \centering
    \begin{subfigure}{0.49\textwidth}
        \includegraphics[width=\textwidth]{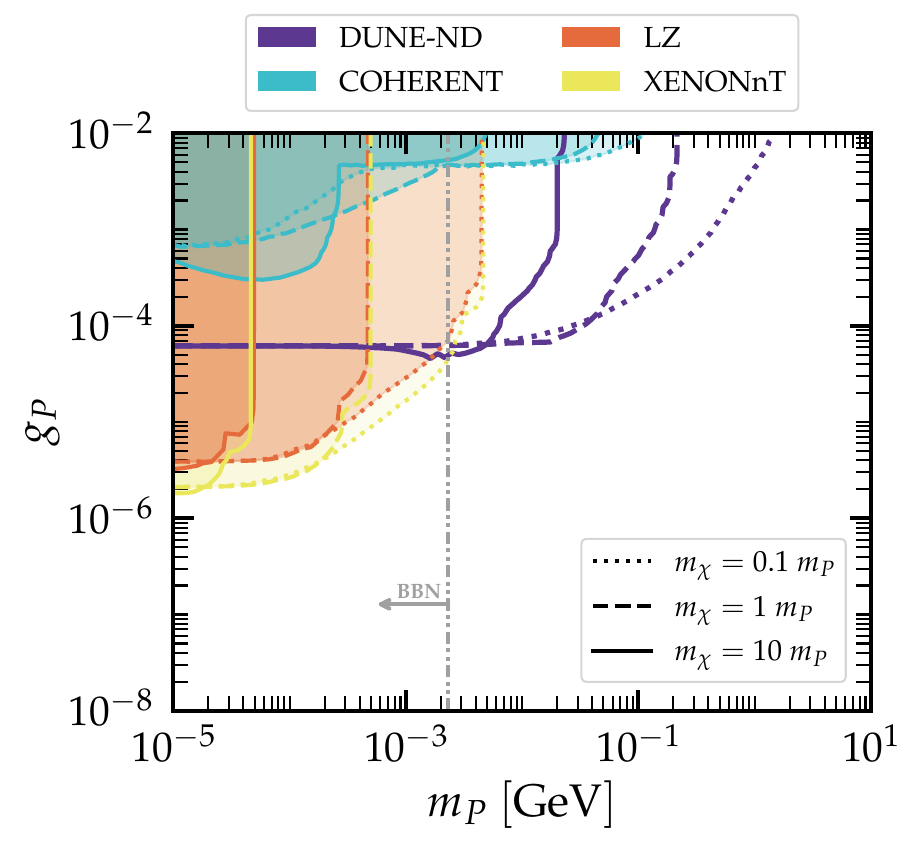}
    \end{subfigure}
    \hfill
    \begin{subfigure}{0.49\textwidth}
        \includegraphics[width=\textwidth]{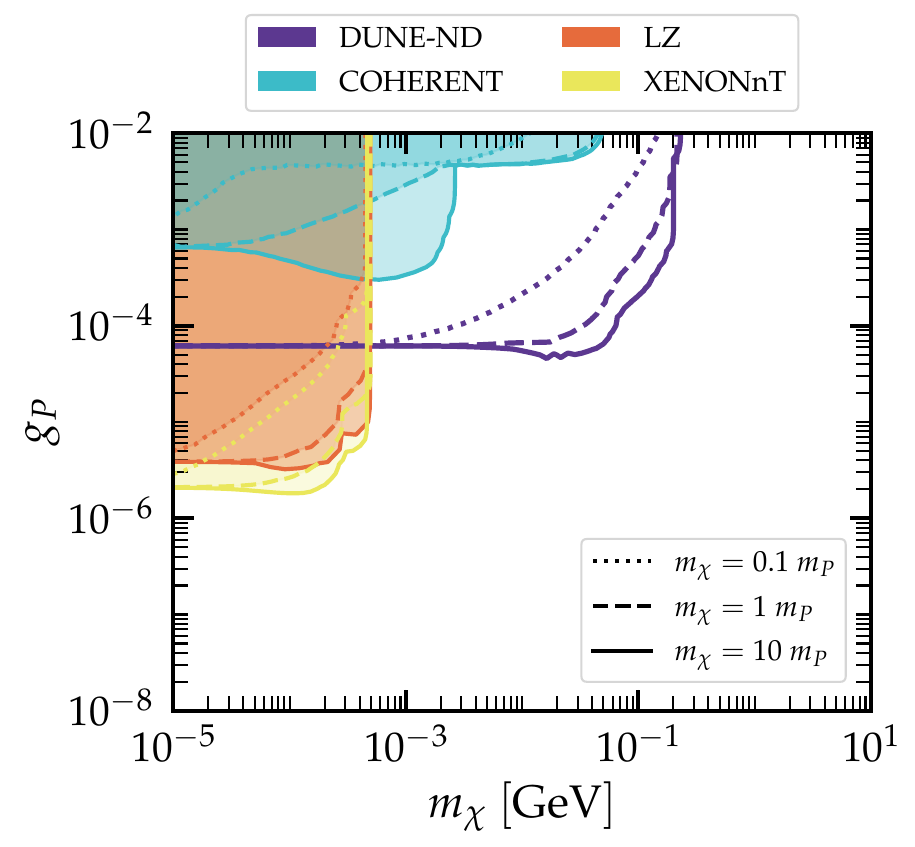}
    \end{subfigure}
    \caption{Same as Fig.~\ref{fig:all-experiments-scalar}, but for  pseudoscalar interactions.}
    \label{fig:all-experiments-pseudoscalar}
\end{figure}

\begin{figure}[!htb]
    \centering
    \begin{subfigure}{0.49\textwidth}
        \includegraphics[width=\textwidth]{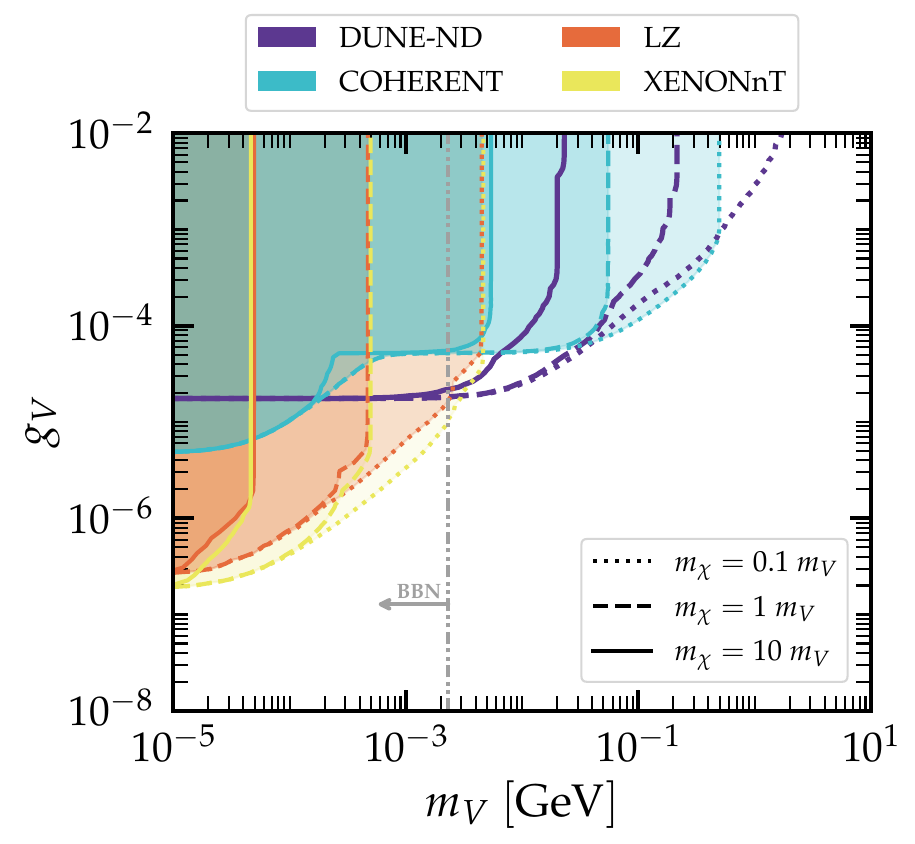}
    \end{subfigure}
    \hfill
    \begin{subfigure}{0.49\textwidth}
        \includegraphics[width=\textwidth]{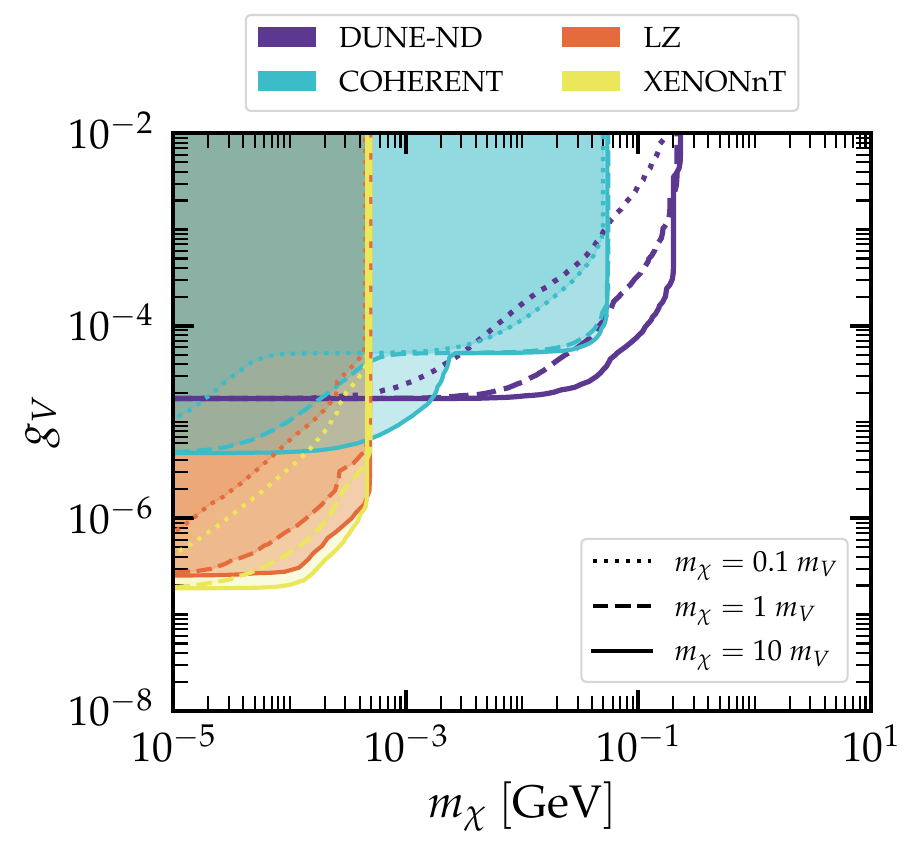}
    \end{subfigure}
    \caption{Same as Fig.~\ref{fig:all-experiments-scalar}, but for  vector interactions.}
    \label{fig:all-experiments-vector}
\end{figure}

\begin{figure}[!htb]
    \centering
    \begin{subfigure}{0.49\textwidth}
        \includegraphics[width=\textwidth]{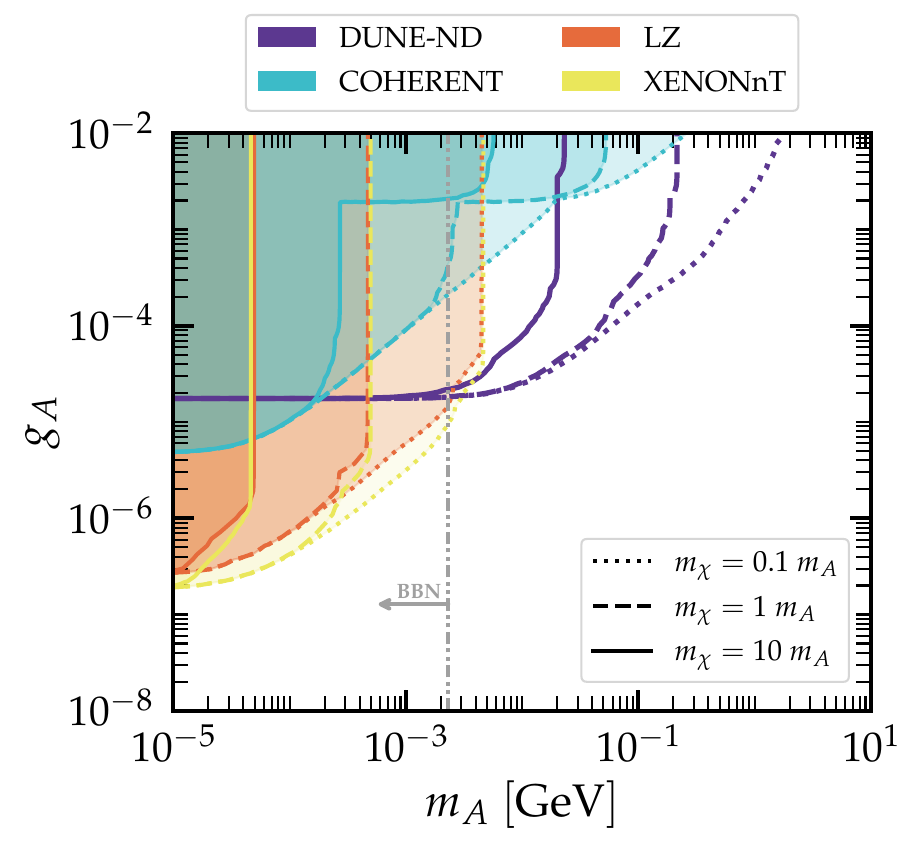}
    \end{subfigure}
    \hfill
    \begin{subfigure}{0.49\textwidth}
        \includegraphics[width=\textwidth]{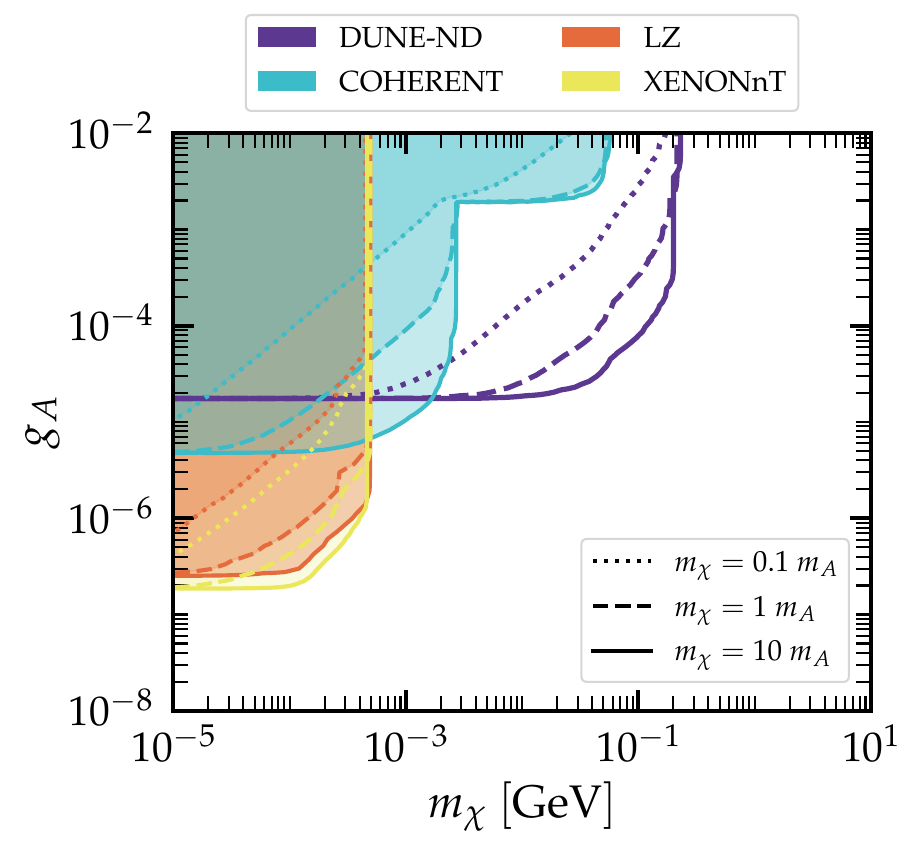}
    \end{subfigure}
    \caption{Same as Fig.~\ref{fig:all-experiments-scalar}, but for  axial-vector interactions.}
    \label{fig:all-experiments-axial}
\end{figure}

\begin{figure}[!htb]
    \centering
    \begin{subfigure}{0.49\textwidth}
        \includegraphics[width=\textwidth]{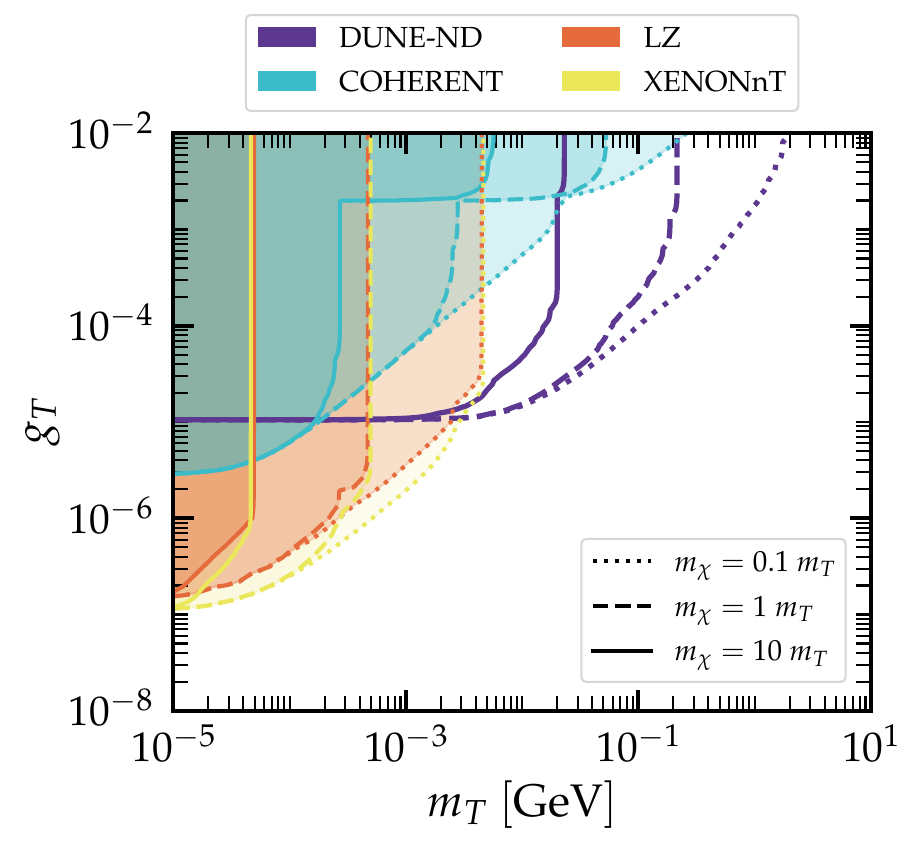}
    \end{subfigure}
    \hfill
    \begin{subfigure}{0.49\textwidth}
        \includegraphics[width=\textwidth]{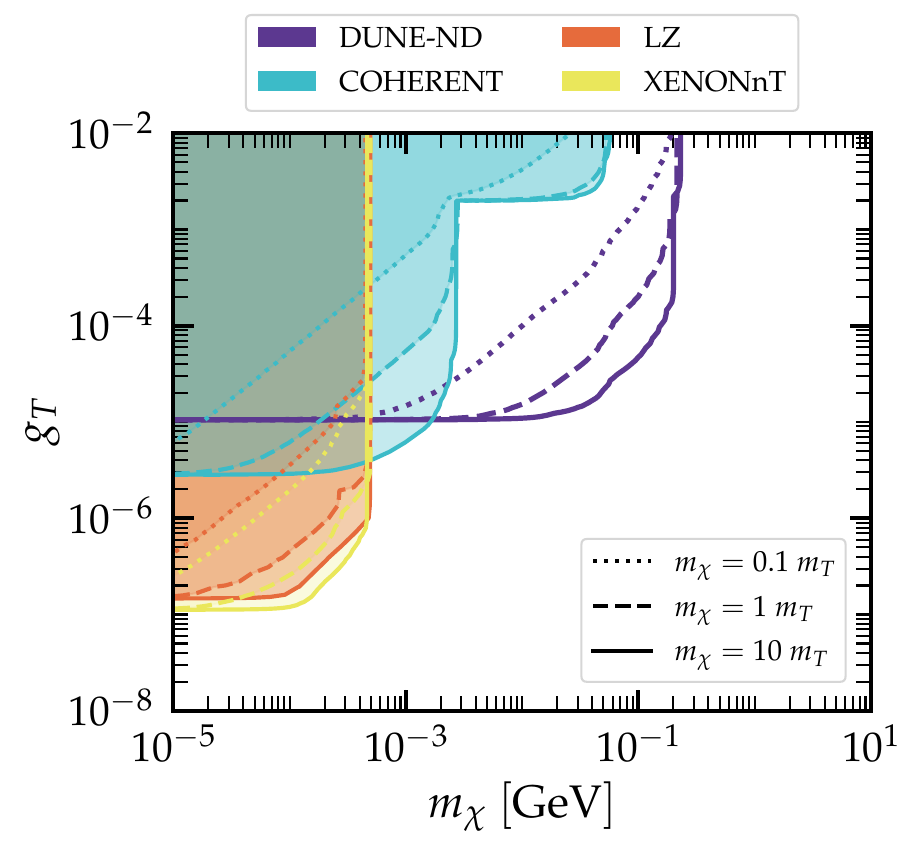}
    \end{subfigure}
    \caption{Same as Fig.~\ref{fig:all-experiments-scalar}, but for  tensor interactions.}
    \label{fig:all-experiments-tensor}
\end{figure}

The results are summarized in Figs.~\ref{fig:all-experiments-scalar}~to~\ref{fig:all-experiments-tensor},  where a comparison of the respective 90\% C.L. sensitivities for scalar, pseudoscalar, vector, axial-vector and tensor interactions, is presented. As before, the analysis has been performed for the three benchmark scenarios $m_\chi= \{0.1, \,1, \,10\} \times \, m_a$. The complementarity of our extracted sensitivities is evident. As can be seen,  the different experiments dominate the sensitivities in different regions of mediator or SF mass. For instance, the XENONnT- and LZ-driven constraints are the most stringent for very low mediator or SF mass. This is due to the very low detection thresholds achievable in LXe DM DD detectors, combined with the inverse recoil-energy dependence in the cross sections. Then, for $m_\chi\geq 300$~keV the recent COHERENT CsI-2021 data are improving the constraints placed by DM DD experiments. At this point, however, it should be noted that the full parameter space probed by DM DD experiments as well as a  large part of it probed with \cevns~data from COHERENT  fall in the region $\lesssim$ few MeV, and may be in conflict with  cosmological and astrophysical observations. The comprehensive estimation of these bounds will eventually depend on the specific model and its thermal history in the early Universe, thus lying beyond the scope of this work. However, we show in Figs.~\ref{fig:all-experiments-scalar},~\ref{fig:all-experiments-pseudoscalar},~\ref{fig:all-experiments-vector}, and ~\ref{fig:all-experiments-axial} a vertical dot-dashed gray line corresponding to a bound imposed with big bang nucleosynthesis (BBN) data (see, e.g.,~\cite{Huang:2017egl}) at the scope of general warning. On the other hand, for sufficiently large mediator or SF mass, i.e.,  $\sim 2$~MeV, the current sensitivities will be improved drastically in view of future measurements at DUNE-ND. This is especially true for the pseudoscalar, axial-vector and tensor interactions, which in the case of \cevns~lead to a spin-dependent  and hence suppressed signal, and  therefore translate into less stringent bounds by COHERENT. Notice however, that COHERENT bounds at $m_\chi \lesssim$ a few MeV are dominated by \eves~events for all interactions except the scalar one, thus leading to competitive limits even in the case of spin-mediated interactions, for which cases the \cevns~contribution is negligible. By exploiting the GeV neutrino beam offered at FNAL it becomes feasible to extend the region of the parameter space that is currently excluded by the COHERENT measurement. Interestingly, in all cases except scalar interactions, DUNE-ND will start dominating the constraints being sensitive to mediator and SF masses in a region not in conflict with cosmological observations.

\subsection{Impact of the \texorpdfstring{$\tau$}{tau}-optimized beam at DUNE-ND}
DUNE will benefit from operating with different horn configurations. So far, we have only assumed the default configuration, which is widely considered in the literature, that is, the \textit{CP-optimized} configuration. At this point, we find it useful to explore the impact of the so-called \textit{$\tau$-optimized} configuration on the SF sensitivities. The latter results into a neutrino energy distribution peaking at higher neutrino energies (see e.g. Refs.~\cite{DUNE:2015lol,Machado:2020yxl}). This in turn will translate into an increase of SF events with larger masses compared to the CP-optimized flux. 

\begin{figure}[!htb]
    \centering
    \begin{subfigure}{0.49\textwidth}
        \includegraphics[width=\textwidth]{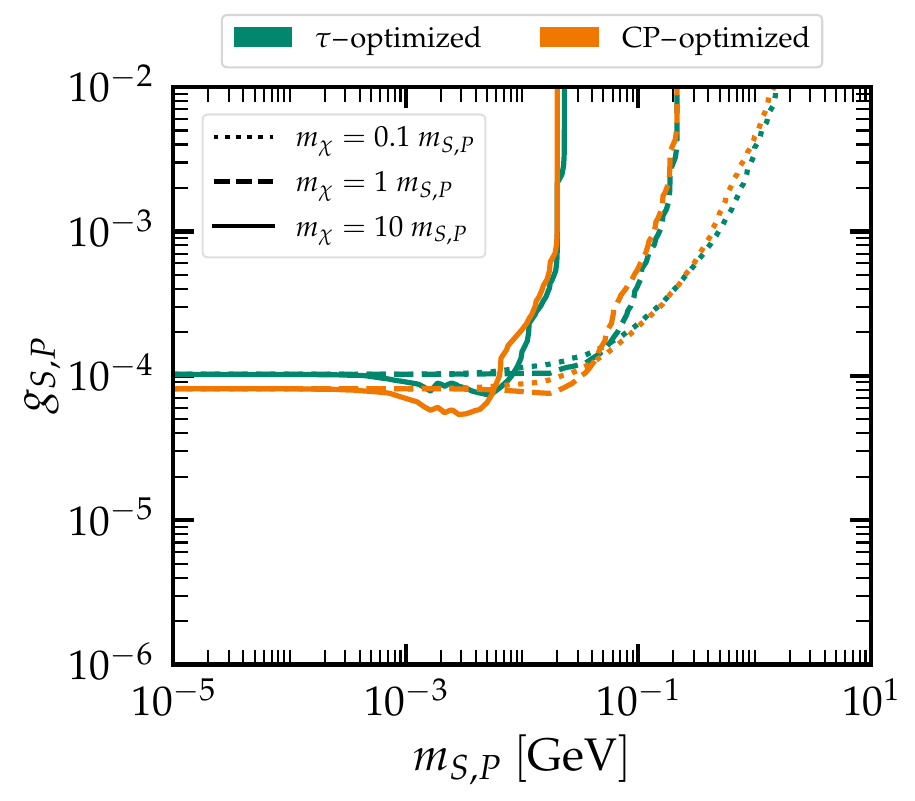}
    \end{subfigure}
    \hfill
    \begin{subfigure}{0.49\textwidth}
        \includegraphics[width=\textwidth]{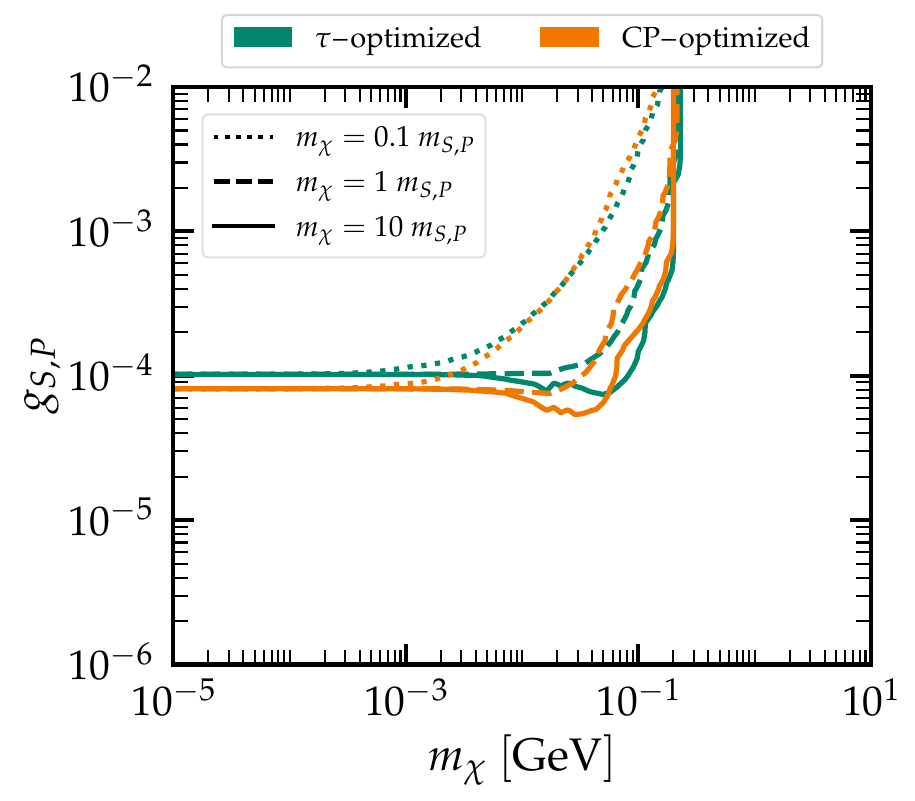}
    \end{subfigure}

    \begin{subfigure}{0.49\textwidth}
        \includegraphics[width=\textwidth]{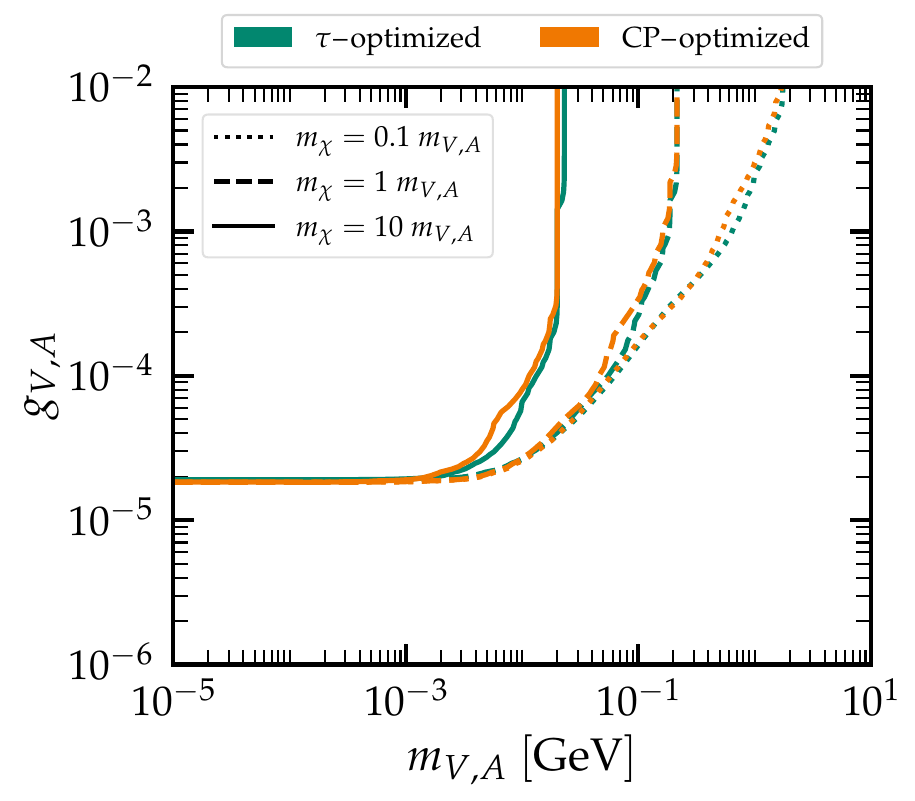}
    \end{subfigure}
    \hfill
    \begin{subfigure}{0.49\textwidth}
        \includegraphics[width=\textwidth]{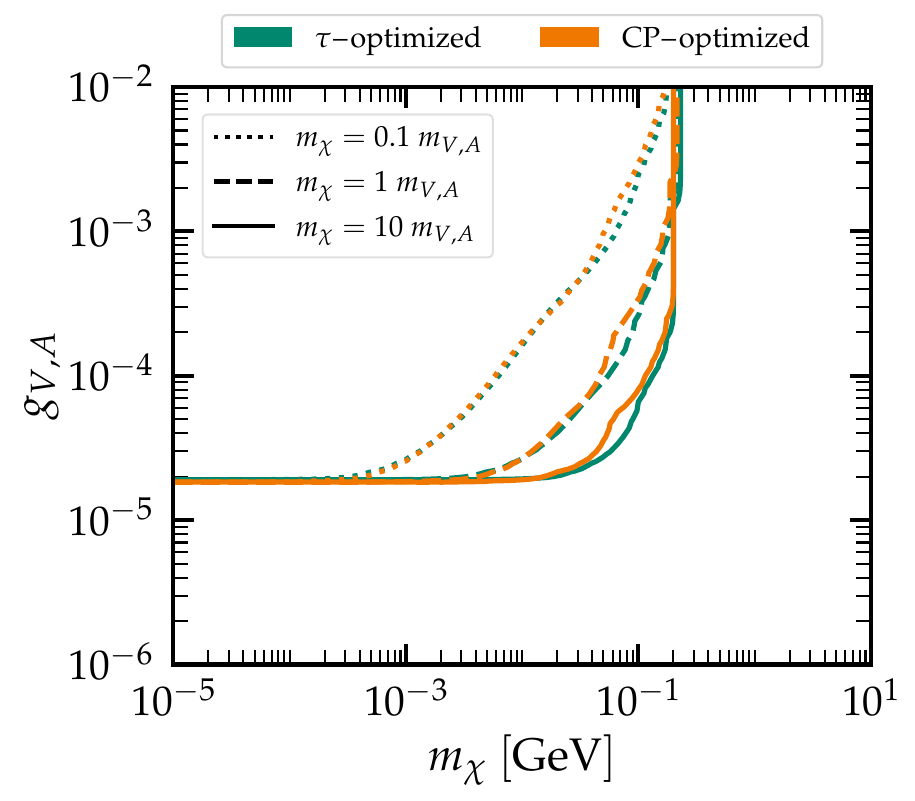}
    \end{subfigure}

    \begin{subfigure}{0.49\textwidth}
        \includegraphics[width=\textwidth]{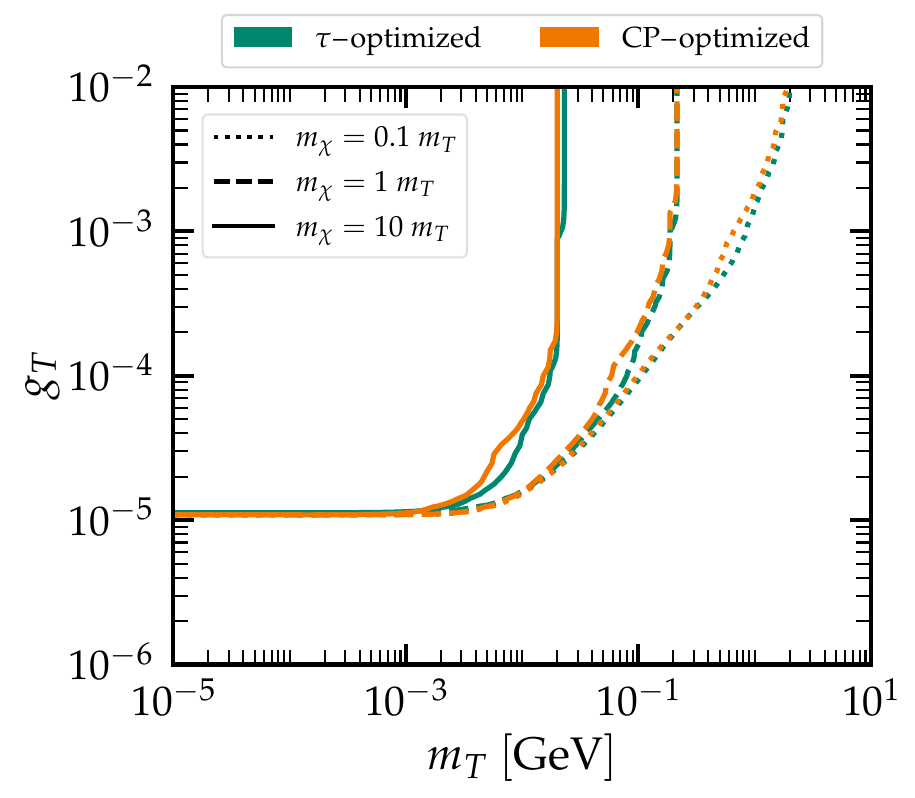}
    \end{subfigure}
    \hfill
    \begin{subfigure}{0.49\textwidth}
        \includegraphics[width=\textwidth]{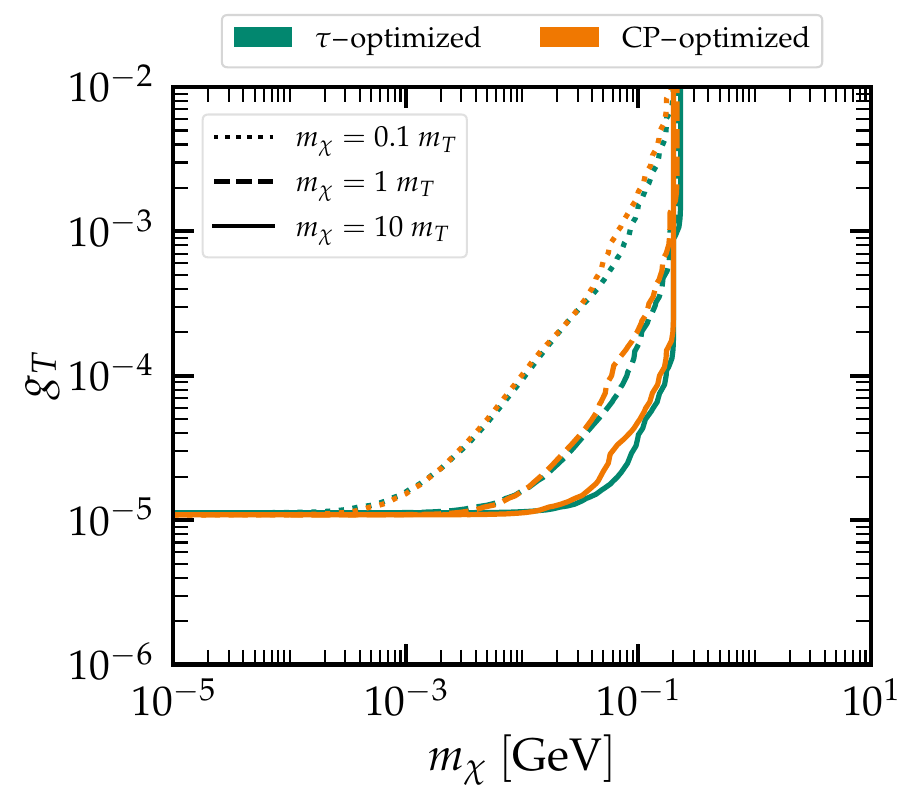}
    \end{subfigure}
    \caption{Comparison of the expected 90\% C.L. sensitivities using the CP-optimized versus the $\tau$-optimized flux at DUNE-ND. The results depict the projected sensitivity assuming 1 yr of exposure for the on-axis location only.}
    \label{fig:CP_vs_tau_opt_flux}
\end{figure}

In Fig.~\ref{fig:CP_vs_tau_opt_flux} we compare the SF sensitivities obtained utilizing the CP- vs. $\tau$-optimized fluxes.  Here, we only consider the on-axis event rates  and 1 year of exposure for both cases. This explains the slight differences of the sensitivities corresponding to the CP-optimized flux shown here vs. Figs.~\ref{fig:dune-locations} and \ref{fig:DUNE_interactions_comparison}. As can be seen from the plots, using the $\tau$-optimized flux leads to slight improvements on the SF sensitivity in the  region of large mediator $m_a$ and $m_\chi$ masses, for the various possible interaction channels. Specially for the scalar/pseudoscalar interaction channel one sees that in the low mass region the CP-optimized flux is more constraining\footnote{This is also the case for the vector/axial-vector and tensor cases as well, but barely visible.}, and depending on the $m_\chi/m_a$ ratio there is eventually a turning point where the constraint driven by the $\tau$-optimized flux starts becoming dominant. Notably, this feature also occurred in heavy sterile neutrino searches in Ref.~\cite{Ovchynnikov:2022rqj}. Finally, recalling the results shown in Figs.~\ref{fig:all-experiments-scalar}--\ref{fig:all-experiments-tensor}, it becomes evident that the comparison of the attainable DUNE-ND sensitivities versus COHERENT and DM DD experiments would have remained essentially unaltered if the $\tau$-optimized flux had been adopted.

\subsection{Sterile fermion decays}

We finally turn our attention to potential signals that could trigger the detector due to events arising from SF decays. Indeed, SFs, after being produced via the up-scattering on nuclei and electrons inside the detector, could eventually decay within the detector to SM particles, thus leading to additional detectable signatures. For the sake of simplicity and illustration, we focus only on the vector interaction. Following Ref.~\cite{Candela:2023rvt}, in our framework there are only two decay modes of $\chi$ at tree-level. Specifically, for $m_\chi > m_V$ ($V$ being the vector mediator) one has the two-body decay $\chi \to V \nu_\ell$, while  for $m_\chi < m_V$ the three-body decay $\chi \to \nu_\ell f\bar{f}$ is possible via an off-shell mediator. 
The  decay widths of the two modes have been previously computed to be~\cite{Jho:2020jfz}~\footnote{Note that in this work we start from an effective Lagrangian (see Eq.~\eqref{eq:effective-lagrangian}). These expressions are obtained in the light-mediator case discussed in Ref.~\cite{Candela:2023rvt} and are equivalent under the assumption $g_V \equiv g_{\chi_L} =  g_f$.}
\begin{align}
    \Gamma(\chi \to V \nu_\ell) &= \dfrac{g_{V}^2}{32\pi} m_\chi \left(1-\dfrac{m_V^2}{m_\chi^2}\right) \left(1 + \dfrac{m_\chi^2}{m_V^2}-2\dfrac{m_V^2}{m_\chi^2}\right), \nonumber \\[4pt]
    \Gamma(\chi \to \nu_\ell \, \ell^- \ell^+) &= \dfrac{g_V^4}{768 \pi^3}\, m_\chi^5 \, I(m_\chi, m_V, \mu_2)\, , \label{eq:chi-3body-decay}
\end{align}
where $I(m_\chi, m_V, \mu_2)$ is expressed as~\cite{Candela:2023rvt}
\begin{align}
    I(m_\chi, m_V, \mu_2) \equiv \int_{0}^{1-4\mu_2} \mathrm{d}x_1 \, &\dfrac{\lambda^{1/2}(1-x_1, \mu_2, \mu_2)}{\left[m_\chi^2(1-x_1) - m_V^2\right]^2}\dfrac{x_1^2}{(1-x_1)^3} \bigg\{6(1+2\mu_2) \bigg. \nonumber \\[4pt]
    &\bigg.- x_1\left[15 + \lambda(1-x_1, \mu_2, \mu_2) + 24\mu_2 + 3x_1^2 - 12x_1(1+\mu_2)\right]\bigg\} \, ,
\end{align}
with $\mu_2 \equiv m_\ell^2 / m_\chi^2$ and $\lambda(x,y,z)=x^2+y^2+z^2-2xy-2xz-2yz$ being the K\"all\'en function. Equation~\eqref{eq:chi-3body-decay} is valid in both the heavy- and light-mediator regimes.  Notice that the latter decay mode is kinematically open for  $m_\chi > 2 m_\ell$~\footnote{As explained in Ref.~\cite{Candela:2023rvt} radiative decays such as $\chi \to \nu \gamma$, $\chi \to \nu \gamma \gamma \gamma$ or $\chi \to \nu \nu \nu$ are possible. However, here we are interested in the analysis of nuclear and electron recoil signals.}.
Finally, if also $m_V>2 m_\ell$, the dilepton decay channel is open and its decay width reads
\begin{equation}
    \Gamma(V \to \ell^- \ell^+) =\frac{g_V^2}{12 \pi } m_V\left(1 + 2 \frac{
   m_\ell^2}{m_V^2}\right)  \sqrt{1-4\frac{m_\ell^2}{m_V^2}} \, .
\end{equation}

 Notice that decay channels involving mesons could also be allowed, depending on the SF mass. While not kinematically accessible at LZ, XENONnT or COHERENT, the decay mode with a $\pi^0$ in the final state would be --- in principle --- allowed at DUNE-ND, but only in a marginal corner of parameter space. For this reason, we did not consider it and we instead focus our discussion on the decays given in Eq.~\eqref{eq:chi-3body-decay}.

\begin{figure}
    \centering
        \begin{subfigure}{0.49\textwidth}
        \includegraphics[width=\textwidth]{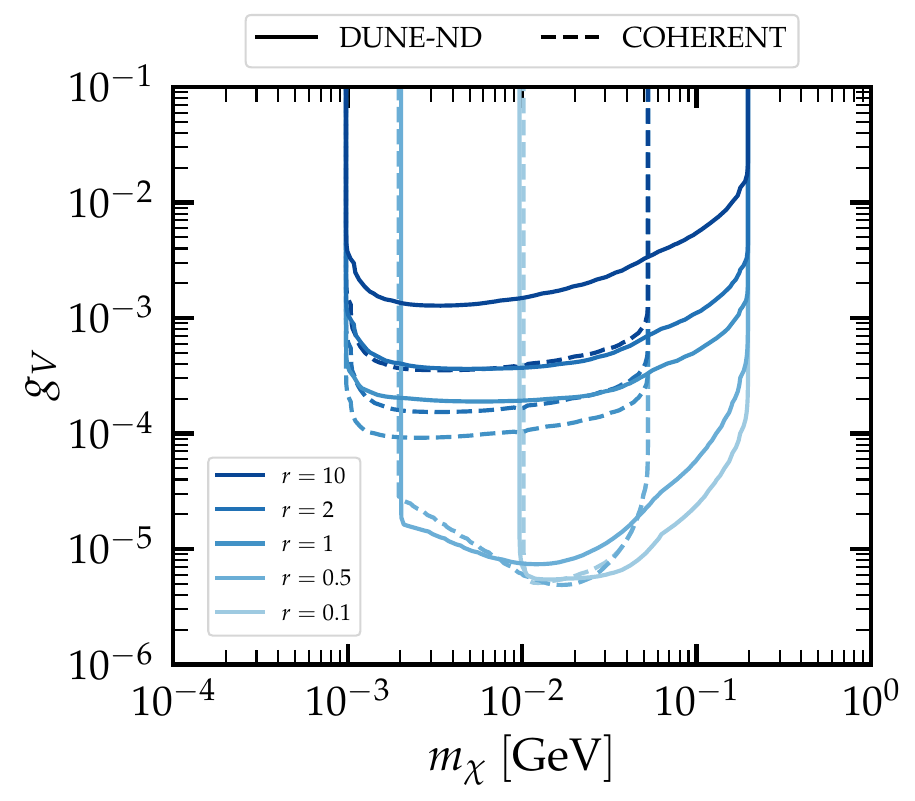}
    \end{subfigure}
    \hfill
    \begin{subfigure}{0.49\textwidth}
        \includegraphics[width=\textwidth]{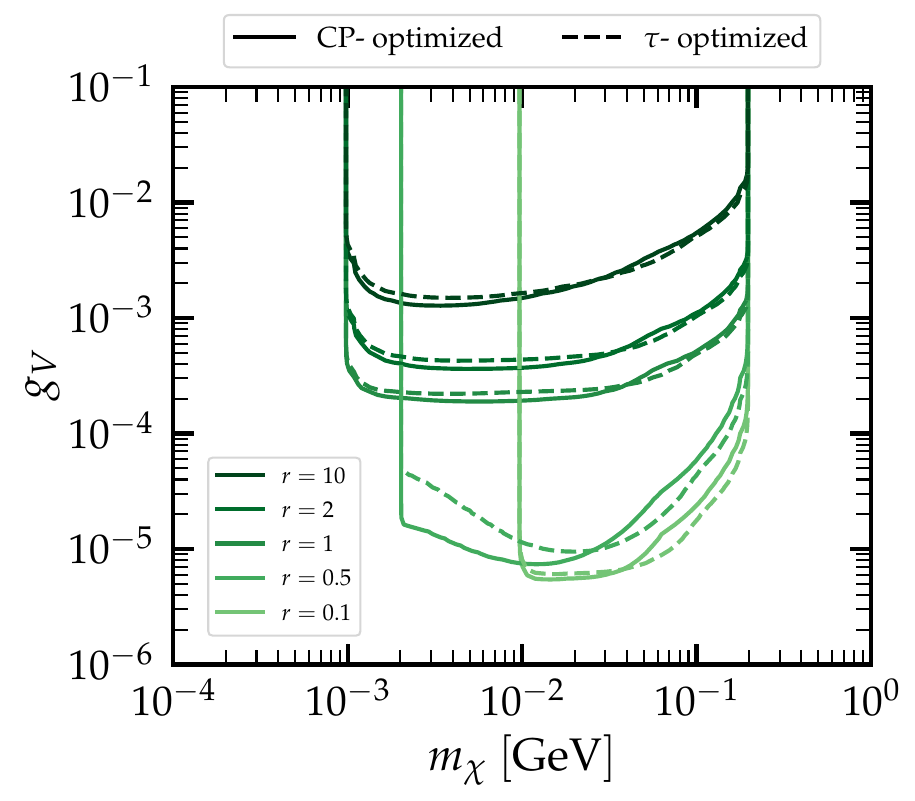}
    \end{subfigure}
    \caption{Left: comparison of the 90\% C.L. exclusion regions extracted from SF decays at COHERENT and DUNE-ND. Right: effect of the CP- versus $\tau$-optimized flux on the sensitivity at DUNE-ND assuming 1 yr of exposure and  the on-axis position. In both figures $r \equiv m_\chi / m_V$.}
        \label{fig:decays}
\end{figure}

We estimate the number of decay events inside the detector, as 
\begin{equation}
    N_\mathrm{decay} = N_0 \int \langle \frac{\mathrm{d} \sigma}{\mathrm{d}E_\chi} \Big\vert^V_\mathrm{\xi} \rangle  P_\mathrm{decay} \, \mathcal{B}(\chi \to ...) \, \epsilon_E \, \, \mathrm{d}E_\chi \,,
    \label{eq:decay_events}
\end{equation}
where  $N_0$  incorporates the neutrino flux normalization, the number of targets and the total time of exposure for a given experiment, while $\mathcal{B}(\chi \to ...) $ stands for the branching ratio of the process. Moreover, $\langle \frac{\mathrm{d} \sigma}{\mathrm{d}E_\chi} \big\vert^V_\mathrm{\xi} \rangle$ represents the differential cross section with respect to the energy of the SF, averaged over the neutrino flux and $\xi=$ 
\{\cevns,~\eves \}. In the previous expression, $\epsilon_E$ denotes the efficiency of electron recoil reconstruction. We may assume that the final state electrons and positrons have on average the same energy, $T_e \sim m_\chi/2$. As a consequence, for the COHERENT experiment the efficiency has a negligible impact. For DUNE-ND, due to the much higher statistics expected at on-axis compared to off-axis locations, only the former is taken into account, assuming an ideal efficiency in order to compute the contours shown in Fig.~\ref{fig:decays}. However, we have checked that our present results are not affected significantly even when a flat, conservative, 15\% efficiency is assumed in the range $50<T_e<m_\chi^\mathrm{max}/2\simeq150$~MeV. The probability for $m_\chi$ to decay inside the detector is defined as  $P_\mathrm{decay}= 1-\exp[-L_\mathrm{det}\Gamma/(\beta \gamma)]$, with $L_\mathrm{det}$ being the detector length along the beam direction (see Ref.~\cite{Ovchynnikov:2022rqj}) and $\beta \gamma$ being the boost factor  to go from the center of mass to the laboratory reference frame.
For the DUNE-ND LArTPC detector we set $L_\mathrm{det}=4$~m~\cite{DUNE:2020fgq}, while for the COHERENT CsI  detector we assume $L_\mathrm{det}=11$~cm~\cite{Konovalov:talk}.

In our analysis, when two-body decays are involved, we furthermore assume that the produced mediator $V$ decays promptly. We do so by introducing a second probability factor in Eq.~(\ref{eq:decay_events}) and requiring that the sum of the two decay lengths does not exceed the size of the detector.

A few comments are in order. First, given the SF masses, $m_\chi$, accessible at DM DD experiments, only the two-body decay $\chi \to V \nu_\ell$ is kinematically allowed. However,  a subsequent decay of $V \to e^- e^+$ is not possible since the up-scattered $m_\chi$ at XENONnT and LZ are always smaller than $2m_e$. Hence, we conclude that SF decays are not expected to leave any extra electronic signal at DM DD experiments.  Next, concerning COHERENT, both the two-body and three-body decays leading to a final state $e^-e^+$ pair are kinematically open, as explained in~\cite{Candela:2023rvt}. Finally, for the case of DUNE-ND two-body and three-body decays that lead to either $e^-e^+$ or $\mu^- \mu^+$ final state pairs are possible, while decays to mesons are ignored. However, the decays to $\mu^- \mu^+$ pairs would modify only slightly the right end of the sensitivity contours since kinematics constrain $m_\chi$ to be up to 200~MeV. We thus set $\mathcal{B}(\chi \to ...) =1$.

Before proceeding to the discussion of our main results, let us stress that, here, our goal is a simplistic order of magnitude estimation,  a detailed Monte Carlo analysis being beyond the scope of our present study. To this purpose, we evaluate the 90\% C.L. sensitivity contours within a background-free framework that is based only on the number of decay events, i.e., we require $N_\mathrm{decay} \geq 2.3$ per year. Finally, in our calculation the decaying particles are assumed to be in the forward direction, which is nonetheless a reasonable assumption as explained in Ref.~\cite{Ovchynnikov:2022rqj}. 

Our results are summarized in Fig.~\ref{fig:decays} where the various 90\% C.L. sensitivity contours correspond to different values of the ratio $r=m_V/m_\chi$. In the left panel of the figure we illustrate a comparison of the exclusion regions  for the cases of COHERENT and DUNE-ND experiments. We observe that DUNE-ND has the prospect to probe a larger region of the parameter space, covering SF masses up to about 200~MeV, while in the case of COHERENT the sensitivity is lost at $\sim$52~MeV, as expected, given the nature of the neutrino source. It is interesting, however, to note the impact of the different detection channels, i.e. neutrino-nucleus vs. neutrino-electron scattering together with the typical energies involved in the two experiments. The COHERENT experiment is exploiting the SNS neutrino beam for which the endpoint occurs at $m_\mu/2\simeq 52.8$~MeV, while in the case of DUNE the FNAL beam endpoint is peaking at  $\sim$2--3~GeV. (The FNAL neutrino beam extends up to 125~GeV, however the relevant part of the neutrino spectrum lies at energies $E_\nu \leq 40$~GeV.) The COHERENT experiment is optimized to detect nuclear scatterings of the heavy CsI target. As a consequence, the recoil energies involved are tiny, and thus most of the incident neutrino energy is available to generate SFs\footnote{The conservation of energy implies $E_\nu = E_\chi - T_\mathcal{N}$, with the nuclear recoil energy being tiny.}. On the other hand, for the case of DUNE where the signal is due to scattering on electrons, most of the incident neutrino energy is carried away through electron recoils, leaving a small amount of energy for SF production. This explains that while GeV-order neutrino energies are involved at DUNE, SFs with a maximum energy of only up to 200~MeV can be probed.

In the right panel of the figure we explore the effect of the obtained sensitivity when the $\tau$-optimized flux is assumed. By comparing the endpoint of the contours where a sharp loss of sensitivity occurs, one can immediately conclude that by using the $\tau$-optimized horn it becomes feasible to probe slightly larger SF masses $m_\chi$. We furthermore find that there is a crossing point for the CP- vs. $\tau$-optimized driven contours. Notice that the same behavior of the contours was found for the scattering-based analysis discussed previously in Fig.~\ref{fig:CP_vs_tau_opt_flux} as well as in the analysis of the dipole portal performed in Ref.~\cite{Ovchynnikov:2022rqj}.

Before closing this discussion we find it useful to elaborate on the shape of the obtained contours. For the cases with $r>1$ (or, equivalently, when $m_V>m_\chi$) that correspond to events coming from the three-body decay $\chi \to \nu_\ell e^- e^+$, the left leg of the contours starts at exactly $m_\chi = 2 m_e$, regardless  of the specific value of $r$. On the other hand, cases with $r<1$ that follow from the two-body decay $\chi \to V \nu_\ell$ and the subsequent decay of the vector mediator $V \to \nu_\ell e^- e+ $, depend upon the condition $m_V \geq 2 m_e$. Hence the left leg of the respective contours depends on the value of the ratio $r$. 

\section{Conclusions}
\label{sec:conc}
In this work we have addressed the possible up-scattering of neutrinos into a new, sterile fermion with a mass in the MeV range. We have considered a simplified phenomenological scenario that allows for all possible Lorentz-invariant non-derivative interactions (scalar,
pseudoscalar, vector, axial-vector and tensor) of neutrinos with electrons and first-generation
quarks.

Motivated by the rich physics opportunities offered by the next-generation long-baseline neutrino oscillation experiment DUNE, we have explored the sensitivity of the DUNE near detector to this scenario. We have simulated \eves~data considering both the nominal neutrino mode and a high-energy configuration. We have found that, while the $\tau$-optimized configuration allows to reach slightly larger sterile fermion masses, the standard configuration actually leads to better sensitivities for the scalar and pseudoscalar interactions. For all other types of interactions, the two configurations produce similar results. 
Additionally, we have investigated how sensitivities change when moving the DUNE near detector off-axis with respect to the neutrino beam. Although going off-axis allows to reduce backgrounds, as expected the neutrino flux is also reduced  and hence the signal, in turn translating  into poorer sensitivities the farther the detector is moved, for all interactions. Eventually, the best projected sensitivity is obtained when the DUNE near detector is on-axis and independently of the beam configuration, allowing to reach couplings as small as $g_{a} \sim 10^{-4},~2 \times 10^{-5},~10^{-5}$ for $a=$ scalar/pseudoscalar, vector/axial-vector and tensor interactions  respectively, and $m_\chi \lesssim$ few/tens of MeV, after 7~years of data taking. The possible decay of the sterile fermion within the same detector may provide additional bounds on the parameter space.

Next, we have compared DUNE sensitivities to bounds obtained with existing data. In this scope, we have analyzed \eves~data from solar neutrinos at current DM DD experiments, LZ and XENONnT and \cevns~data from the most recent result of the COHERENT-CsI experiment. We have found that all these facilities dominate the sensitivities in different regions of mediator or sterile fermion mass. Namely, XENONnT and LZ dominate the constraints at very small masses; COHERENT bounds exclude smaller couplings in the intermediate ($\sim$ tens of MeV) mass range, while DUNE definitely will provide the most stringent constraints at heavier masses, being able to probe sterile fermions as heavy as $\sim 200$~MeV. These considerations hold especially for spin-dependent  interactions, for which \cevns~data are suppressed. We have finally explored the possibility of detecting additional signatures from sterile fermion decays inside the detectors under consideration. While such decays are not relevant for DM DD experiments, for COHERENT and DUNE-ND our results lead to the same general conclusions as in the case of up-scattering.

To conclude, our work is meant to show the complementarity of independent facilities involving different neutrino sources to a phenomenological scenario in which a neutrino can up-scatter into a sterile fermion. As a follow-up, it would be interesting to pursue further theoretical work to explore realistic UV-completions of this phenomenological framework.

\section*{Acknowledgments}
This work has been supported by the Spanish grants PID2020-113775GB-I00 (MCIN/AEI/ 10.13039/501100011033) and CNS2023-144124 (MCIN/AEI/10.13039/501100011033 and “Next Generation EU”/PRTR). VDR acknowledges financial support by the CIDEXG/2022/20 grant (project ``D'AMAGAT'') funded by Generalitat Valenciana. PMC is supported by the grant CIACIF/2021/281 also funded by Generalitat Valenciana. The work of PM, DKP and NS was supported by the Hellenic Foundation for Research and Innovation (H.F.R.I.) under the “3rd Call for H.F.R.I. Research Projects to support Post-Doctoral Researchers” (Project Number: 7036).

\noindent

\appendix

\section{Hadronic physics}
\label{sec:hadronic-physics}

The couplings $C_a$ defined in Eqs.~\eqref{eq:couplingsCa}-\eqref{eq:couplingsCV} depend upon the quark mass and spin contributions to the nucleons: $f_{T_q}^{(p)}$ and $f_{T_q}^{(n)}$ represent the quark mass contributions to the nucleon (proton and neutron) mass; $\Delta_{q}^{(p)}$ and $\Delta_{q}^{(n)}$ parametrize the quark spin content of the nucleon; and $\delta_{q}^{(p)}$ and $\delta_{q}^{(n)}$ are tensor charges that characterize the difference between the spin of quarks and anti-quarks inside the nucleon.  Even though the determination of some of these parameters is still uncertain, in this work we fix the following values \cite{DelNobile:2021wmp}
\begin{align}
    f_{T_u}^{(p)} &= 0.026 , & f_{T_u}^{(n)} &= 0.018 \,, \\[4pt]
    f_{T_d}^{(p)} &= 0.038, & f_{T_d}^{(n)} &= 0.056 \,, \\[4pt]
    \Delta_{u}^{(p)} &= \Delta_{d}^{(n)} = 0.777, & \Delta_{d}^{(p)} &= \Delta_{u}^{(n)} = -0.438 \,, \\[4pt]
    \delta_{u}^{(p)} &= \delta_{d}^{(n)} = 0.784, & \delta_{d}^{(p)} &= \delta_{u}^{(n)} = -0.204\,.
\end{align}
Note that for the pseudoscalar and tensor parameters (last two rows) the isospin symmetry limit has been assumed, i.e., $\delta_{u}^{(p)} = \delta_{d}^{(n)}$, $\delta_{d}^{(p)} = \delta_{u}^{(n)}$, $\Delta_{u}^{(p)} = \Delta_{d}^{(n)}$ and $\Delta_{d}^{(p)} = \Delta_{u}^{(n)}$. Finally, let us comment that using different values~\cite{Belanger:2013oya,Anselmino:2008jk} for these charges would imply, at most, a difference of $27\%$ in the scalar and tensor couplings, of $10\%$ in the axial-vector coupling $g_A$ and of $12\%$ in the pseudoscalar coupling $g_P$.

\section{Spin structure functions and cross sections}
\label{app:SpinSfunc}
Nuclear spin-dependent  effects are incorporated via the spin structure function $S^\kappa(\qtransfer^2)$, defined as
\begin{equation}
    S^\kappa(\qtransfer^2) = (g_X^0)^2 S^\kappa_{00} + g_X^0 g_X^1 S^\kappa_{01} + (g_X^1)^2 S^\kappa_{11} \, ,
\end{equation} 
where $X = A,\, T$ denotes the axial-vector or tensor interaction respectively, and $\kappa = \mathcal{L},\, \mathcal{T}$ represents the longitudinal ($\mathcal{L}$) or transversal ($\mathcal{T}$) contribution from nuclear operators.
The explicit form of $S^{\kappa}_{ij}$ is given in~\cite{Hoferichter:2020osn} and can receive both contributions\footnote{In this work we neglect two-body current corrections (see Ref.~\cite{Hoferichter:2020osn} for details).}. Here, $g_X^0$ and $g_X^1$ stand for the isoscalar and isovector couplings, respectively, which take the form
\begin{equation}
    g_X^0 = \dfrac{g_X^p + g_X^n}{2}, \qquad g_X^1 = \dfrac{g_X^p - g_X^n}{2} \, .
    \label{eq:iso_couplings}
\end{equation}

\subsection{Axial-vector interaction}
To compute the axial-vector cross section we adopt the multipole expansion of the hadronic current given by Donnelly-Walecka ~\cite{Donnelly:1976fs,Donnelly:1978tz}. By keeping only the dominant multipoles and ignoring vector/axial-vector interference terms, one has
\begin{equation}
   \left.\dfrac{\d \sigma_{\nu_\ell \mathcal{N}}}{\d T_\mathcal{N}}\right|_\mathrm{CE \nu NS}^\mathrm{A} (E_\nu, T_\mathcal{N}) = \dfrac{m_\mathcal{N}}{2 J + 1} \dfrac{E_\chi}{E_\nu}\left[ 2 l_3 l_3^*~S^\mathcal{L}(\qtransfer^2) + \left( \vec{l} \cdot \vec{l} -l_3 l_3^* \right)~S^\mathcal{T}(\qtransfer^2) \right] \, ,
\end{equation}
where the lepton traces are defined as
\begin{equation}
    \sum_\mathrm{spins} l^{\mu \nu} = \frac{1}{4 E_\nu E_\chi} \mathrm{Tr}\left[ (\slashed{k'} + m_\chi) \gamma^\mu P_L \slashed{k}  \gamma^\nu P_L \right] \, ,
\end{equation}
with $k'$ and $k$ being the 4-momentum of the upscattered sterile fermion and of the incoming neutrino, respectively.
We note that for the case of a neutrino upscattering into a sterile fermion, we find that both the longitudinal and the transverse contributions are relevant in contrast to the axial-vector SM CE$\nu$NS. This is due to the fact that the respective leptonic trace $l_3l_3^*$  is not suppressed as in the SM case~\cite{Hoferichter:2020osn}.  The proton $(g_A^p)$ and neutron $(g_A^n)$ couplings are obtained through the expression
\begin{equation}
    g_A^N =  \sum_{q = u,d}  g_q^A \Delta_q^{(N)} = g_f \sum \Delta_q^{(N)}\,,  \qquad N=p,n \, ,
\end{equation}
where in the second equality we have assumed universal quark couplings i.e., $g_u^A=g_d^A=g_f$. Then, Eq.~(\ref{eq:iso_couplings}) for $X=A$ reads
\begin{equation}
\begin{aligned}
    g_A^0 &= g_f \dfrac{\Delta_u^{(p)} + \Delta_d^{(p)} + \Delta_u^{(n)} + \Delta_d^{(n)}}{2} \equiv g_f \tilde{g}_A^0 \,, \\[4pt]
    g_A^1 &= g_f \dfrac{\Delta_u^{(p)} + \Delta_d^{(p)} - \Delta_u^{(n)} - \Delta_d^{(n)}}{2} \equiv g_f \tilde{g}_A^1 \, ,
\end{aligned}
\end{equation}
and the corresponding axial-vector structure finally takes the form
\begin{equation}
    S^{\kappa}(\qtransfer^2) = g_f^2 \left[(\tilde{g}_A^0)^2 S_{00}^{\kappa} + \tilde{g}_A^0 \tilde{g}_A^1 S_{01}^{\kappa} + (\tilde{g}_A^1)^2 S_{11}^{\kappa}\right] \equiv g_f^2 \tilde{S}^{\kappa}(\qtransfer^2)\, .
\end{equation}
A direct comparison with the axial-vector component of the Lagrangian in Eq.~(\ref{eq:effective-lagrangian}) yields
\begin{equation}
    g_f = \dfrac{G_F}{\sqrt{2}} \varepsilon_\ell^A \,,
\end{equation}
which in the light-mediator regime finally becomes
\begin{equation}
    g_f^2 = \dfrac{G_F^2}{2} (\varepsilon_\ell^A)^2 \longrightarrow \dfrac{g_A^4}{(m_A^2 + \qtransfer^2)^2}\, .
\end{equation}

\subsection{Tensor  interaction}

For tensor interactions the multipole expansion of the hadronic current is given in Eq.~(D2) of Ref.~\cite{Hoferichter:2020osn} (see also Ref.~\cite{Glick-Magid:2022erc}). Following the discussion of Ref.~\cite{Hoferichter:2020osn} and keeping the dominant terms, the resulted cross section reads
\begin{equation}
   \left.\dfrac{\d \sigma_{\nu_\ell \mathcal{N}}}{\d T_\mathcal{N}}\right|_\mathrm{CE \nu NS}^\mathrm{T} (E_\nu, T_\mathcal{N}) = \dfrac{m_\mathcal{N}}{2 J + 1} \dfrac{E_\chi}{E_\nu}\left[ 2 l_3^{(1)} l_3^{(1)*}~S^\mathcal{L}(\qtransfer^2) + \left( \vec{l^{(1)}} \cdot \vec{l^{(1)}} -l_3^{(1)} l_3^{(1)*} \right)~S^\mathcal{T}(\qtransfer^2) \right] \, .
\end{equation}
The corresponding leptonic traces are given by~\cite{Hoferichter:2020osn} 
\begin{equation}
    \sum_{\mathrm{spins}} l_i^{(1)} l_j^{(1)} = \epsilon_{ikl} \epsilon_{jmn} L_{klmn}  \, ,
\end{equation}
with 
\begin{equation}
   L^{\mu \nu \lambda \sigma} = \frac{1}{4 E_\nu E_\chi} \mathrm{Tr}\left[ (\slashed{k'} + m_\chi) \sigma^{\mu \nu} P_L \slashed{k}  \sigma^{\lambda \sigma}  P_R \right] \, .
\end{equation}
Analogously to the axial-vector case, the spin structure function for tensor interactions is written as
\begin{equation}
    S^\kappa(\qtransfer^2) = (g_{T}^0)^2 S_{00}^\kappa + g_{T}^0 g_{T}^1 S_{01}^\kappa + (g_{T}^1)^2 S_{11}^\kappa,
\end{equation}
with $\kappa = \mathcal{T},\, \mathcal{L}$ for the transversal and longitudinal component, respectively. The corresponding isoscalar and isovector couplings are given by
\begin{equation}
    g_{T}^0 = \dfrac{g_{T}^p + g_{T}^n}{2}, \qquad g_{T}^1 = \dfrac{g_{T}^p - g_{T}^n}{2} \, ,
\end{equation}
where for universal quark couplings one gets
\begin{equation}
    g_{T}^N = \sum_{q = u,d} g_q^T \delta_q^{(N)} = g_f \sum_{q = u,d}  \delta_q^{(N)}\, .
\end{equation}
Working in a similar manner as for the axial vector case, we extract the dependence on $g_f$ explicitly by defining
\begin{equation}
\begin{aligned}
    g_{T}^0 &= g_f \dfrac{\delta_u^{(p)} + \delta_d^{(p)} + \delta_u^{(n)} + \delta_d^{(n)}}{2} \equiv g_f \tilde{g}_{T}^0, \\[4pt]
    g_{T}^1 &= g_f \dfrac{\delta_u^{(p)} + \delta_d^{(p)} - \delta_u^{(n)} - \delta_d^{(n)}}{2} \equiv g_f \tilde{g}_{T}^1.
\end{aligned}
\end{equation}
Hence, the structure function for tensor interactions reads,
\begin{equation}
    S^\kappa(\qtransfer^2) = g_f^2 \left[(\tilde{g}_{T}^0)^2 S_{00}^\kappa + \tilde{g}_{T}^0 \tilde{g}_{T}^1 S_{01}^\kappa + (\tilde{g}_{T}^1)^2 S_{11}^\kappa\right] \equiv g_f^2 \tilde{S}^\kappa(\qtransfer^2).
\end{equation}
As before, comparing with the tensor component of the Lagrangian in Eq.~(\ref{eq:effective-lagrangian}), one finds
\begin{equation}
    g_f = \dfrac{G_F}{\sqrt{2}} \varepsilon_\ell^T \, ,
\end{equation}
which in the light-mediator regime eventually gives
\begin{equation}
    g_f^2 = \dfrac{G_F^2}{2} (\varepsilon_\ell^T)^2 \longrightarrow \dfrac{g_T^4}{(m_T^2 + \qtransfer^2)^2}\, .
\end{equation}

\section{Effective electron charge for Cs, I and Xe}\label{sec:appendix-effective-charge}
In this Appendix we provide the effective electron charge values for Cs, I and Xe (see Table~\ref{tab:ZeffCsI}).
\begin{table}[!htb]
\centering
    \begin{tabular}{cc@{ $< T_e \leq$ }c}
    \toprule
    $\boldsymbol{Z^\mathrm{Cs}_\mathrm{eff}}$ & \multicolumn{2}{c}{$\boldsymbol{T_e~(}$\textbf{keV}$\mathbf{)}$} \\
    \midrule
    55 & \multicolumn{2}{c}{$T_e >$ 35.99} \\[2pt]
    53 & 5.71 & 35.99 \\[2pt]
    51 & 5.36 & 5.71 \\[2pt]
    49 & 5.01 & 5.36 \\[2pt]
    45 & 1.21 & 5.01 \\[2pt]
    43 & 1.07 & 1.21 \\[2pt]
    41 & 1.00 & 1.07 \\[2pt]
    37 & 0.74 & 1.00 \\[2pt]
    33 & 0.73 & 0.74 \\[2pt]
    27 & 0.23 & 0.73 \\[2pt]
    25 & 0.17 & 0.23 \\[2pt]
    23 & 0.16 & 0.17 \\[2pt]
    19 & \multicolumn{2}{c}{$T_e \leq$ 0.16} \\
    \bottomrule
    \end{tabular}
    \hfil
    \begin{tabular}{cc@{ $< T_e \leq$ }c}
    \toprule
    $\boldsymbol{Z^\mathrm{I}_\mathrm{eff}}$ & \multicolumn{2}{c}{$\boldsymbol{T_e~(}$\textbf{keV}$\mathbf{)}$} \\
    \midrule
    53 & \multicolumn{2}{c}{$T_e >$ 33.17} \\[2pt]
    51 & 5.19 & 33.17 \\[2pt]
    49 & 4.86 & 5.19 \\[2pt]
    47 & 4.56 & 4.86 \\[2pt]
    43 & 1.07 & 4.56 \\[2pt]
    41 & 0.93 & 1.07 \\[2pt]
    39 & 0.88 & 0.93 \\[2pt]
    35 & 0.63 & 0.88 \\[2pt]
    31 & 0.62 & 0.63 \\[2pt]
    25 & 0.19 & 0.62 \\[2pt]
    23 & 0.124 & 0.19 \\[2pt]
    21 & 0.123 & 0.124 \\[2pt]
    17 & \multicolumn{2}{c}{$T_e \leq$ 0.123} \\
    \bottomrule
    \end{tabular}
    \hfil
    \begin{tabular}{cc@{ $< T_e \leq$ }c}
    \toprule
    $\boldsymbol{Z^\mathrm{Xe}_\mathrm{eff}}$ & \multicolumn{2}{c}{$\boldsymbol{T_e~(}$\textbf{keV}$\mathbf{)}$} \\
    \midrule
    54 & \multicolumn{2}{c}{$T_e >$ 34.76} \\[2pt]
    52 & 5.51 & 34.76 \\[2pt]
    50 & 5.16 & 5.51 \\[2pt]
    48 & 4.84 & 5.16 \\[2pt]
    44 & 1.17 & 4.84 \\[2pt]
    42 & 1.02 & 1.17 \\[2pt]
    40 & 0.96 & 1.02 \\[2pt]
    36 & 0.71 & 0.96 \\[2pt]
    32 & 0.69 & 0.71 \\[2pt]
    26 & 0.23 & 0.69 \\[2pt]
    24 & 0.18 & 0.23 \\[2pt]
    22 & 0.16 & 0.18 \\[2pt]
    18 & 0.074 & 0.16 \\[2pt]
    14 & 0.072 & 0.074 \\[2pt]
    10 & 0.028 & 0.072 \\[2pt]
    4 & 0.013 & 0.028 \\[2pt]
    2 & 0.012 & 0.013 \\[2pt]
    0 & \multicolumn{2}{c}{$T_e \leq$ 0.012} \\
    \bottomrule
    \end{tabular}
    \caption{\centering{Effective electron charge for Cs, I~\cite{Thompson:booklet} and Xe~\cite{Chen:2016eab} as a function of the energy deposition $T_e$}.}
    \label{tab:ZeffCsI}
\end{table}

\bibliographystyle{utphys}
\bibliography{bibliography}

\end{document}